\documentclass[prb, superscriptaddress,twocolumn,10pt,
 amsmath,
 amssymb,
 aps,
]{revtex4-2}

\usepackage{graphicx}

\usepackage{bm}
\usepackage[dvipsnames]{xcolor}
\usepackage{subfigure}

\usepackage{hyperref}

\newcommand{\beq}[1]{\begin{equation} #1 \end{equation}}

\newcommand{\komutator}[2]{[#1,#2]}

\def\<{\left\langle}
\def\>{\right\rangle}

\newcommand{\vk}{\textbf{k}} 
\newcommand{\vp}{\textbf{p}}

\definecolor{darkpink}{rgb}{0.91, 0.33, 0.5}

\begin{document}

\title{Collective excitations and divergent spin currents in non-centrosymmetric superconductors}

\author{Markus Lysne}
\thanks{Currently at Sintef Energy Research}
\affiliation{Department of Physics, University of Fribourg, 1700 Fribourg, Switzerland}

\author{Philipp Werner}
\affiliation{Department of Physics, University of Fribourg, 1700 Fribourg, Switzerland}

\author{Nikolaj Bittner}
\thanks{Currently at Zeiss SMT}
\affiliation{Department of Physics, University of Fribourg, 1700 Fribourg, Switzerland}

\date{\today}

\begin{abstract}
We study the collective modes in a non-centrosymmetric superconductor with Rashba spin-orbit coupling under laser irradiation. The concept of Anderson Pseudospin Resonance allows to reveal how laser driving gives rise not only to the established resonant enhancement of the third harmonic response, but also to a resonant enhancement in the second harmonic response of the spin current. We propose a theory which explains the phenomenon without including interband transitions. The theory is corroborated by numerical simulations which incorporate interband effects and allow us to clarify the signatures of the collective modes in the long-time dynamics of the superconductor. 
\end{abstract}

\maketitle

\section{Introduction}
 
Systems with spin-orbit coupling exhibit interesting physics such as spin-momentum locking and the spin-Hall effect. A hallmark of these systems is the combination of 
time reversal symmetry and lack of inversion symmetry. These properties were shown by Gor'kov to have implications for the superconducting pairing in the sense that an admixture of spin singlet and spin triplet pairing is allowed \cite{Gorkov_2001} -- an insight which has given rise to the field of non-centrosymmetric superconductors (NCS)~\cite{NCS2012}. Despite numerous theoretical~\cite{frigeri:2004a, frigeri:2004b, frigeri:2004c, Frigeri_2006, Samokhin_2008, Vorontsov_2008, klam:2009, Bittner_2015} and experimental~\cite{bauer:2004, togano:2004, badica:2005} studies of non-centrosymmetric superconductivity in different bulk materials, the detailed characterization of the superconducting gap by means of experimental probes still remains a challenge.

On the other hand, the dynamics of superconductors under light irradiation has in recent years attracted considerable interest. The application of ultra-short THz light pulses with a photon energy tuned to the superconducting energy gap opened a new path for the study of
superconducting condensates using nonequilibrium probes. In particular, time-resolved experiments enabled the direct detection of collective modes~\cite{matsunaga:2013, giorgianni:2019} and the study of symmetry properties of the superconducting gap~\cite{chu:2020}. On the theoretical side, the signatures of the amplitude (Higgs) mode of the superconducting order parameter in the third harmonic generation (THG) response have been clarified in the pioneering work by Tsuji {\it et al.} \cite{Tsuji_2015}. Following these studies, several investigations of nonequilibrium superconductors with unconventional pairing and competing order parameters were conducted \cite{Bittner_2015, Murotani_2017, Muller_2018, Schwarz_2020, Muller_2021, fiore:2022}. However, while in the last years the collective modes have been extensively studied in conventional superconducting
systems with a spin-singlet order parameter, as well as some unconventional superconductors, less is known about the phenomenon in NCS superconductors. The role of Higgs and Leggett (phase) modes in THG in non-centrosymmetric systems has only recently been investigated, using diagrammatic techniques~\cite{Klein_2024}.

\begin{figure} 
  \includegraphics[width=0.85\linewidth]{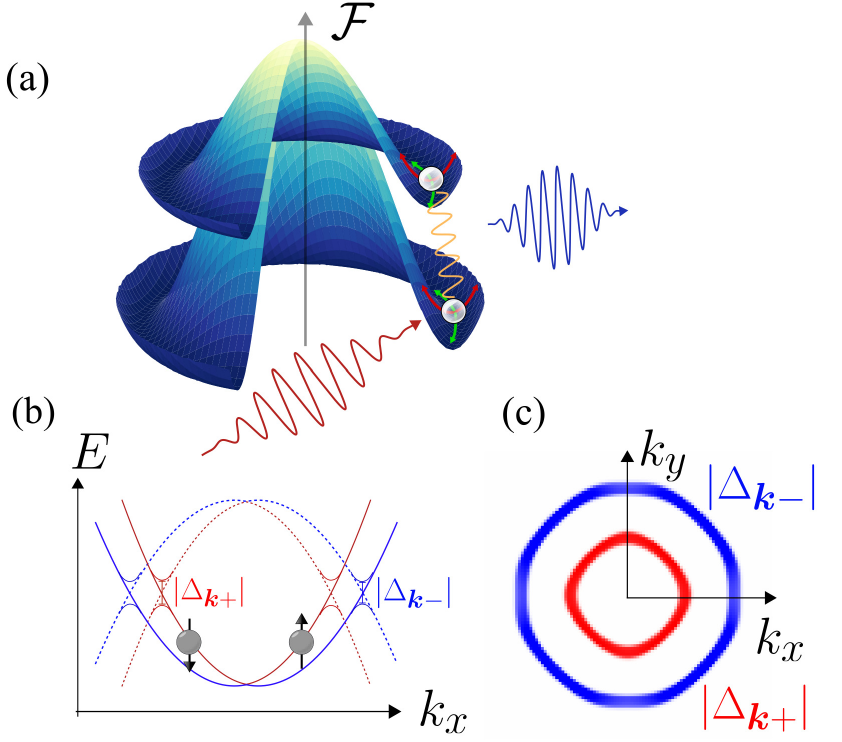} 
  \caption{(a) Free-energy landscape of the system with the order parameters for the two coupled condensates indicated by spheres. Amplitude and phase modes, as well as Leggett modes are indicated by red and green arrows, respectively, while the Leggett mode is represented by an orange wavy line connecting the two condensate order parameters.
  The system's nonlinear light-matter response is illustrated by the  field pulses. (b) The Rashba model dispersion presented in particle-hole space along with the anti-crossing resulting from the superconducting pairing. (c) Illustration of the order parameter values on the different Fermi surfaces.
  }
  \label{fig:mexhat}
\end{figure}

In this work, we study the light-matter response of a special class of NCS superconductors, namely, systems with Rashba SOC, focusing on the interplay between collective excitations and the resonant enhancement of spin currents. The collective behavior of the system admits a description similar to that of two-band superconductors \cite{Krull_2016}, as illustrated in Fig.~\ref{fig:mexhat}(a). The band structure of the system is shown in (b) and its two-component superconducting condensate as illustrated in (c). Beyond the well-known possibility of harboring mixed singlet-triplet superconducting pairing, the absence of inversion symmetry gives rise to rich spin dynamics. While previous studies (e.g., Ref.~\cite{Silaev_2020}) explored spin current responses in specific superconducting configurations, here we demonstrate that photo-induced spin currents emerge as a generic feature in systems with SOC. We will demonstrate that nonlinear phenomena, such as the second-harmonic generation (SHG) response of the spin current, can be resonantly enhanced by the collective mode dynamics, following a similar mechanism as that of Ref.~\cite{Tsuji_2015}. In order to explain said phenomena, we outline a formalism for studying collective modes, as well as nonlinear responses in different physical observables. 

The paper is organized as follows. In Section~\ref{sec:Formalism} we introduce the model and our theoretical methods. Section~\ref{sec:Results} shows results of our numerical and analytical calculations for the system in and out of equilibrium. Finally, we summarize our work in Section~\ref{sec:Summary}.

\section{Formalism}
\label{sec:Formalism}

\subsection{Hamiltonian}

We start by introducing the Hamiltonian describing a two-dimensional interacting non-centrosymmetric (NCS) superconductor, $\hat{H}=\hat{H}_{0} + \hat{H}_{I}$, as the sum of a non-interacting part, $\hat{H}_{0}$, and an interacting part $\hat{H}_{I}$. 
The non-interacting part of the
Hamiltonian can be expressed as 
\begin{equation} \label{eq:SOC}
\begin{aligned}
	\hat{H}_0 &= \sum_{\vk, s_{1}s_{2}} h_{\vk; s_{1},s_{2}} \hat{c}_{\vk s_{1}}^\dagger \hat{c}_{\vk s_{2}} \\
		   &= \sum_{\vk,s_{1}s_{2}} \big[(\epsilon_{\vk}^0-\mu_{c}) \tau_{0} + \boldsymbol{d}_{\vk} \cdot \boldsymbol{\tau}\big]_{s_{1},s_{2}} \hat{c}_{\vk s_{1}}^\dagger \hat{c}_{\vk s_{2}} ,
\end{aligned}
\end{equation}
where $\hat{c}_{\vk s}$ is the annihilation operator for a Fermion with momentum $\vk$ and spin $s$, $\epsilon_{\vk}^0$ is the electron dispersion without spin-orbit coupling, $\mu_{c}$ the chemical potential, $\boldsymbol{\tau}$ a vector of Pauli matrices, and $\boldsymbol{d}_{\vk}$ represents the anti-symmetric spin-orbit coupling. $H_0$ can be diagonalized by a unitary transformation $U_{\vk}$, specified in appendix \ref{app:interaction}, which results in the Hamiltonian 
\begin{equation}\label{eq:nonintham}
	\hat{H}_{0} = \sum_{\vk\mu} \epsilon_{\vk\mu} \hat{a}_{\vk \mu}^\dagger \hat{a}_{\vk \mu},
\end{equation}
with $\epsilon_{\vk\mu} = \epsilon^0_{\vk} + \mu |\boldsymbol{d}_{\vk}| - \mu_{c}$. Here $\hat{a}_{\vk \mu}$ is the annihilation operator for an electron with momentum $\vk$ and band index $\mu=\pm$. The role of $\boldsymbol{d}_{\vk}$ is to lift the band degeneracy between the two states at a given momentum. As we study the Rashba Hamiltonian with the Rashba spin-orbit coupling parameter $\alpha$, we introduce the vector $\boldsymbol{g}_{\vk}$, defined through $\boldsymbol{d}_{\vk} = \alpha \boldsymbol{g}_{\vk}$. 

The interacting part of the Hamiltonian can be expressed as \cite{Smidman_2017} 
\begin{equation} \label{eq:int}
	\hat{H}_{I} = \frac{1}{2}\sum_{\vk \vk'}\sum_{s_{1}s_{2}; s_{1}'s_{2}'} V_{\vk,\vk'; s_{1}s_{2} s_{2}'s_{1}'} \hat{c}_{\vk s_{1}}^\dagger \hat{c}_{-\vk s_{2}}^\dagger \hat{c}_{-\vk' s_{2}'} \hat{c}_{\vk' s_{1}'} ,
\end{equation}
where the interaction strength can be represented in the following form~\cite{Frigeri_2006, Samokhin_2008} 
\begin{equation} \label{eq:interactionspinbasis}
\begin{aligned}
	V_{\vk,\vk'; s_{1}s_{2}s_{2}'s_{1}'} =& \frac{V_{0}}{4} \Big\{ e_{s} \phi_{s_{1},s_{2}} \phi_{s_{2}',s_{1}'}^\dagger \\
	& + e_{{t}} (\boldsymbol{g}_{\vk}\cdot \boldsymbol{\phi})_{s_{1},s_{2}}(\boldsymbol{g}_{\vk'}\cdot \boldsymbol{\phi})_{s_{2}',s_{1}'}^\dagger  \\
	& + e_{{m}} (\boldsymbol{g}_{\vk}\cdot \boldsymbol{\phi})_{s_{1},s_{2}}\phi_{s_{2}',s_{1}'}^\dagger \\
	&+ e_{{m}} \phi_{s_{1},s_{2}} (\boldsymbol{g}_{\vk'}\cdot \boldsymbol{\phi})_{s_{2}',s_{1}'}^\dagger  \Big\} .
\end{aligned}
\end{equation}
Here, $V_{0}\leq 0$ parametrizes the interaction strength, and $e_s$, $e_{tr}$, and $e_m$ are non-negative parameters corresponding to the singlet, triplet and mixed interaction strengths, respectively. They fulfill the relation 
\begin{equation}
e_s^2 + e_{tr}^2 + e_{m}^2 = 1. 
\end{equation}
We have also introduced the matrices $\phi = i\tau_{y}$ and $\boldsymbol{\phi}=i\boldsymbol{\tau}\tau_{y}$. The interaction terms involving $e_{m}$ originate from the Dzyaloshinskii-Moriya interaction~\cite{dzyaloshinskii:1958, moriya:1960, Frigeri_2006}. 

With Eq.~\eqref{eq:int} expressed in the band basis, we make the approximation of only retaining intraband pairing contributions. The resulting interaction term reads
\begin{equation} \label{eq:helicity}
	\hat{H}_{I} \approx \frac{1}{2}\sum_{\vk \vk'}\sum_{\mu\nu} V_{\vk,\vk'; \mu\nu} \hat{a}_{\vk \mu}^\dagger \hat{a}_{-\vk \mu}^\dagger \hat{a}_{-\vk' \nu} \hat{a}_{\vk' \nu},
\end{equation}
with
\beq{
\label{eq:IntStrgth}
\begin{aligned}
V_{\vk\vk';\mu\nu} =& \frac{V_{0}}{4} \bigg[ e_{s} R_{\vk; \mu\mu}^s R_{\vk';\nu\nu}^{s,\dagger} + e_{tr}R_{\vk;\mu\mu}^{tr} R_{\vk';\nu\nu}^{tr,\dagger}  \\
	&+ e_{m} R_{\vk;\mu\mu}^{tr} R_{\vk' ; \nu\nu}^{s,\dagger} + e_{m} R_{\vk; \mu\mu}^{s} R_{\vk';\nu\nu}^{tr,\dagger} \bigg]. 
\end{aligned}
}
Here, $R_{\vk}^s  = U_{\vk}^\dagger i\tau_{y} U_{-\vk}^*$ and $ R_{\vk}^{tr} = U_{\vk}^\dagger (i \boldsymbol{g}_{\vk}\cdot\boldsymbol{\tau}) \tau_{y} U_{-\vk}^* $. Equation~\eqref{eq:IntStrgth} reduces to a separable form for either pure singlet or pure triplet pairing. A derivation of this expression can be found in Appendix \ref{app:interaction}. 

For future reference, let us note that the interaction can be re-expressed as
\begin{equation} \label{eq:int_band_basis}
\begin{aligned}
	V_{\vk\vk'; \mu\nu} =& \frac{V_{0} e^{ -i (\phi_{\vk} - \phi_{\vk'})}}{4} \big[ \mu\nu e_{s} + e_{tr}  f_{\vk} f_{\vk'} \\
	& \hspace{22mm} + e_{m}(\nu f_{\vk} + \mu f_{\vk'}) \big],
\end{aligned}
\end{equation}
where we have introduced an even and real function, $f_{\vk}$, as defined in Appendix \ref{app:interaction} and $\phi_{\vk} = \tan^{-1}(d_{\vk,y}/d_{\vk,x})$. Here $\mu,\nu=\pm 1$. The formula holds under the assumption of $d_{z}=0$.

Upon performing a mean-field decoupling of Eq.~\eqref{eq:helicity}, we can write (up to a constant energy shift)
\begin{equation} \label{eq:mf_decoupling}
	\hat{H}_{I} \approx  \frac{1}{2}\sum_{\vk\mu } \big( \Delta_{\vk\mu} \hat{a}_{\vk \mu}^\dagger \hat{a}_{-\vk \mu}^\dagger + \Delta_{\vk\mu}^{*} \hat{a}_{-\vk \mu} \hat{a}_{\vk \mu} \big),
\end{equation}
where the band-selective order parameters are
\beq{
\label{eq:EnGapEq}
\Delta_{\vk\mu} = \sum_{\vp\nu} V_{\vk\vp;\mu\nu}\<\hat a_{-\vp\nu}\hat a_{\vp\nu}\>,
}
with $ \langle \dots \rangle \equiv \mathrm{Tr}[ \hat{\rho} \dots ]$ and $\hat{\rho}$ the density matrix of the system. For simplicity, we adopt the notation $\sum_{\vk\nu} \mbox{``="} N^{-1} \sum_{\vk\nu} $ with $N$ the number of ${\bf  k}$-points in the first Brillouin Zone (BZ), which we use throughout for momentum sums of expectation values. Additionally, the (normalized) BZ sum in Eq.~\eqref{eq:EnGapEq} is restricted to a region around the Fermi surfaces with an energy width corresponding to a given Debye frequency, $\omega_{D}$. 

In appendix \ref{app:interaction}, we derive the order parameter arising from this interaction, which is given by
\begin{equation} \label{eq:orderparmexpression}
	\Delta_{\vk\mu} = -\mu e^{-i\phi_{\vk}} (\Delta_{s} + \mu f_{\vk} \Delta_{tr}),  
\end{equation}
where $\Delta_{s}$, $\Delta_{tr}$ are the singlet and triplet contributions, respectively, and the phase $\phi_{\vk}$ fulfills $e^{i\phi_{-\vk}} = - e^{i\phi_{\vk}}$, giving the symmetry $\Delta_{-\vk\mu} = -\Delta_{\vk\mu}$ required by the Fermion anti-commutation relations and Eq.~\eqref{eq:mf_decoupling}. Our expression coincides with Eq. (40) in Ref.~\cite{Smidman_2017}.

\subsection{Anderson pseudospin representation} \label{sec:B}

In this section, we express the mean-field Hamiltonian (Eqs.~\eqref{eq:nonintham}+\eqref{eq:mf_decoupling}) in the Anderson pseudospin (APS) language, which is convenient for studying collective modes \cite{Vorontsov_2008}. Using the Pauli matrices $\tau^{x,y,z}$, as well as the identity matrix $\tau^0$, we may write
\beq{
\label{eq:MFH}
\hat{H}_\text{APS}^\text{eq} = \sum_{\vk\mu}\boldsymbol{\hat{\sigma}}_{\vk\mu}\cdot\boldsymbol{b}_{\vk\mu},
}
where we have used the definition of the Anderson pseudospin operators \cite{Tsuji_2015},  
\beq{ \label{eq:Pauli_operators}
\hat{\sigma}_{\vk\mu}^{i} = \frac{1}{2} \begin{pmatrix} \hat{a}_{\vk,\mu}^\dagger & \hat{a}_{-\vk,\mu} \end{pmatrix}\cdot\tau^{i}\cdot  \begin{pmatrix} \hat{a}_{\vk,\mu} \\ \hat{a}_{-\vk,\mu}^\dagger \end{pmatrix},
}
with $i = \{x, y, z,0\}$ and where 
$$
\boldsymbol{b}_{\vk\mu} = 
\left(
\begin{array}{ccc}
       \Delta'_{\vk\mu}, &
       -\Delta''_{\vk\mu}, &
      \frac{1}{2}\left(\epsilon_{\vk\mu} + \epsilon_{-\vk\mu}\right)
\end{array}
\right)
$$
can be interpreted as a pseudomagnetic field acting on the Anderson pseudospin vector
$\boldsymbol\sigma_{\vk\mu} \equiv \langle \boldsymbol{\hat{\sigma}}_{\vk\mu} \rangle$. Here, $\epsilon_{\vk\mu}$ is defined below Eq.~\eqref{eq:nonintham} and $\Delta'_{\vk\mu}$ represents the real and $\Delta''_{\vk\mu}$ the imaginary part of the order parameter.
Analogously, Eq.~\eqref{eq:EnGapEq} can be recast into the form
\beq{
\label{eq:selfcons}
\Delta_{\vk\mu} = \sum_{\vp\nu} V_{\vk\vp; \mu\nu} \left[\sigma_{\vp\nu}^x - i\sigma_{\vp\nu}^y\right].
}

We further assume that the electron system is exposed to an electric field pulse, which can be incorporated into the Hamiltonian by means of the Peierls substitution \cite{Schuler_2021}. This leads to a time-dependent modulation of the $z$-component of the pseudomagnetic field, $b_{\vk\mu}^z (t)$, whose form is derived in Appendix~\ref{app:perturbation}. The electric field perturbation, expressed in terms of the vector potential $\boldsymbol{A}(t)$ with $\boldsymbol{E}(t) = -\partial_{t} \boldsymbol{A}(t)$, is polarized in the plane of the sample and is assumed to be spatially homogeneous. The effect on the model is given by 
\begin{equation} \label{eq:bz}
	{b}_{\vk\mu}^z(t) = \frac{1}{2}( \tilde{\epsilon}_{\vk,q\boldsymbol{A}(t),\mu} + \tilde\epsilon_{-\vk,q\boldsymbol{A}(t),\mu}),
\end{equation}
where we have defined 
$$
	\tilde{\epsilon}_{\vk, \boldsymbol{p},\mu} = [U_{\vk}^\dagger h_{\vk - \boldsymbol{p}} U_{\vk}]_{\mu\mu}
$$ 
with the electronic charge, $q=-|e|$, defined in terms of the elementary charge, $e$. Equation~\eqref{eq:bz} follows from a rather subtle rewriting of $\sum_{\vk\mu} \tilde\epsilon_{\vk,q\boldsymbol{A}(t),\mu} \hat{a}_{\vk \mu}^\dagger \hat{a}_{\vk \mu}$ according to Eq.~\eqref{eq:gen_obs} in which the term proportional to $\hat{\sigma}_{\vk\mu}^{0}$ is discarded as $[\hat{\sigma}_{\vk\mu}^{0},\hat{\sigma}_{\vk\mu}^{i}]=0$ for $i=x,y,z$. While previous studies showed a cancellation of the linear-in-$A$ term \cite{Tsuji_2015, Murotani_2017}, this is not necessarily the case in our formulation, as this depends on the nature of $h_{\vk}$ and the transformation, $U_{\vk}$.

Denoting the resulting time-dependent Hamiltonian by $\hat{H}_\text{APS}(t)$, the time-dependent Anderson pseudospin vector obeys the (Bloch) equation of motion, 
\begin{align}
&\partial_t\boldsymbol{\hat{\sigma}}_{\vk\mu}(t) = i\komutator{\hat{H}_{APS}(t)}{\boldsymbol{\hat{\sigma}}_{\vk\mu}(t)}  \nonumber\\
&= \big( {b}_{\vk\mu}^x(t) -  {b}_{-\vk\mu}^x (t), {b}_{\vk\mu}^y(t) -  {b}_{-\vk\mu}^y (t), {b}_{\vk\mu}^z(t) +  {b}_{-\vk\mu}^z (t) \big) \nonumber\\
& \hspace{4mm}\times{\hat{\sigma}}_{\vk\mu}(t)  \nonumber\\
&= 2\big( {b}_{\vk\mu}^x(t), {b}_{\vk\mu}^y(t) , {b}_{\vk\mu}^z(t) \big) \times{\hat{\sigma}}_{\vk\mu}(t),\label{eq:EoM}
\end{align}
where we have used 
\begin{equation} \label{eq:commutators}
	[\hat{\sigma}_{\vk\mu}^{i}, \hat{\sigma}_{\vk\mu}^{j}] = i\epsilon_{ijk} \hat{\sigma}_{\vk\mu}^{k},
\end{equation}
with $\epsilon_{ijk}$ the Levi-Civita tensor and where the second line follows from the symmetries $\hat{\sigma}_{-\vk\mu}^{i} = -\hat{\sigma}_{\vk\mu}^{i}$ for $i = \{x, y\}$ and $\hat{\sigma}_{-\vk\mu}^{z} = \hat{\sigma}_{\vk\mu}^{z}$. The last equality in \eqref{eq:EoM} follows from the symmetries $\Delta_{-\vk\mu} = -\Delta_{\vk\mu}$ and ${b}_{\vk\mu}^z(t) = {b}_{-\vk\mu}^z(t)$. Equation~\eqref{eq:EoM} must be solved self-consistently 
together with the time-dependent energy gap equation~\eqref{eq:selfcons}. 

We emphasize that in contrast to the case studied in e.g.~\cite{Tsuji_2015}, the standard commutation relations of Eq.~\eqref{eq:commutators} hold provided that we are not in one of the high symmetry points of the BZ (for which $\vk = -\vk$ modulo a reciprocal lattice vector). This issue can be addressed by using a shifted $\vk$-space discretization which avoids all such points. 

\subsection{Linearized equations of motion}

If we assume that the external electromagnetic field is sufficiently weak, a linearization of Eq.~\eqref{eq:EoM} is possible \cite{Tsuji_2015, Murotani_2017}. To that end, we define
\begin{equation} \label{eq:linearization}
\begin{aligned}
    \Delta_{\vk\mu}(t) &= \Delta_{\vk\mu}(0) + \delta\Delta_{\vk\mu}(t), \\
    \boldsymbol{\sigma}_{\vk\mu}(t) &= \boldsymbol{\sigma}_{\vk\mu}(0) + \delta\boldsymbol{\sigma}_{\vk\mu}(t), \\
    {b}_{\vk\mu}^z(t) &= {b}_{\vk\mu}^z(0) + \delta{b}_{\vk\mu}^z(t), \\
\end{aligned}
\end{equation}
with $\delta\Delta(t)$, $\delta\boldsymbol\sigma(t)$ and $\delta{b}_{\vk\mu}^z(t)$ denoting the deviations from the equilibrium values. The components of $\boldsymbol\sigma_{\vk\mu}(0)$ are computed via a Bogoliubov transformation, 
	$$ \hat{a}_{ \vk\mu} = w_{ \vk\mu}^{*} \hat{\gamma}_{ \vk\mu} + v_{ \vk\mu} \hat{\gamma}_{ -\vk\mu}^\dagger, $$
with $|w_{ \vk\mu}^2  | + |v_{ \vk\mu}^2  |=1$, $w_{ -\vk\mu}=w_{ \vk\mu} $, $v_{ -\vk\mu}=-v_{ \vk\mu} $ and $\hat{\gamma}_{\vk\mu}$ obeying Fermionic anti-commutation relations as defined in Ref.~\cite{Weng_2016}. Specifically, 
$w_{\vk\mu} = \sqrt{\frac{1}{2}(1 + \frac{\epsilon_{\vk\mu}}{E_{\vk\mu}})}$ and $v_{\vk\mu} = -e^{i \mathrm{arg}(\Delta_{\vk\mu})} \sqrt{\frac{1}{2}(1 - \frac{\epsilon_{\vk\mu}}{E_{\vk\mu}})}$. Since the resulting Bogoliubov quasiparticles obey standard Fermi-Dirac statistics (with, e.g., $\langle \hat{\gamma}_{\vk \mu}^\dagger \hat{\gamma}_{\vk \mu} \rangle=n_{F}(E_{\vk \mu})$, $E_{\vk\mu} = \sqrt{\epsilon_{\vk\mu}^2 + |\Delta_{\vk\mu} |^2 }$ and $n_{F}(\epsilon)$ the Fermi-Dirac distribution), we obtain
$$
	\boldsymbol\sigma_{\vk\mu}(0) = - \frac{1}{2 E_{\vk \mu}} \tanh \bigg( \frac{E_{\vk \mu}}{2k_{B} T}\bigg) (\Delta'_{\vk\mu},
       -\Delta''_{\vk\mu}, 
      \epsilon_{\vk\mu} ).
$$
The resulting equations of motion can be derived by inserting Eq.~\eqref{eq:linearization} into Eq.~\eqref{eq:EoM} and discarding quadratic terms in the deviations. After performing a Fourier transform, we get 
\begin{equation} \label{eq:linearize}
  \delta \boldsymbol{\sigma}_{\vk\mu}(\omega) = M_{\vk\mu}(\omega) \cdot \delta \boldsymbol{b}_{\vk\mu}(\omega)
\end{equation}
with
\begin{widetext}
\begin{equation}
\label{eq:BlochFT}
  { M}_{ \vk\mu}(\omega) = \frac{ \tanh(\beta E_{ \vk\mu}/2) }{{ E}_{ \vk \mu} ( \omega^2 - 4{ E}_{ \vk \mu}^{ 2} )} 
  \begin{pmatrix}
    2( { \epsilon}_{ \vk\mu}^{ 2} + { \Delta}_{ \vk\mu}^{ '' 2} ) &  2{ \Delta}_{ \vk\mu}^{ '} { \Delta}_{ \vk\mu}^{ ''} + i { \epsilon}_{ \vk\mu} \omega & - 2{ \epsilon}_{ \vk\mu} { \Delta}_{ \vk\mu}^{ '} + i { \Delta}_{ \vk\mu}^{ ''} \omega \\
     2{ \Delta}_{ \vk\mu}^{ '} { \Delta}_{ \vk\mu}^{ ''} - i { \epsilon}_{ \vk\mu} \omega & 2( { \epsilon}_{ \vk\mu}^{ 2} + { \Delta}_{ \vk\mu}^{ ' 2} ) &  2{ \epsilon}_{ \vk\mu} { \Delta}_{ \vk\mu}^{ ''} + i { \Delta}_{ \vk\mu}^{ '} \omega \\
    - 2{ \epsilon}_{ \vk\mu} { \Delta}_{ \vk\mu}^{ '} - i { \Delta}_{ \vk\mu}^{ ''} \omega &  2{ \epsilon}_{ \vk\mu} { \Delta}_{ \vk\mu}^{ ''}-i { \Delta}_{ \vk\mu}^{ '} \omega &  2( { \Delta}_{ \vk\mu}^{ ' 2} + { \Delta}_{ \vk\mu}^{ '' 2}) \\  
  \end{pmatrix} ,
\end{equation}
\end{widetext}
following the analysis of Ref.~\cite{Muller_2021}. (Note that we have flipped the sign of the imaginary part relative to the convention $ \Delta = \Delta^{'} - i \Delta^{''}$ in Ref.~\cite{Muller_2021}.) \\

In what follows, we will derive a closed set of equations for various components of the order parameter as well as an expression for the susceptibility which will provide more insight into the frequency of the collective modes. We begin by considering the deviation from equilibrium of Eq.~\eqref{eq:selfcons} in the Fourier domain, which reads
\begin{equation} \label{eq:general_delta_OP}
\delta\Delta_{\vk\mu}(\omega) = \sum_{\vp\nu} V_{\vk\vp;\mu\nu} \left[\delta\sigma_{\vp\nu}^x(\omega) - i\delta\sigma_{\vp\nu}^y(\omega)\right].
\end{equation}
We also define the deviations
\beq{
\label{eq:EnGap}
\begin{split}
   \delta\tilde\Delta'_{\vk\mu}(\omega) = \sum_{\vp\nu} V_{\vk\vp;\mu\nu} \delta\sigma_{\vp\nu}^x(\omega),\\
   \delta\tilde\Delta''_{\vk\mu}(\omega) = - \sum_{\vp\nu} V_{\vk\vp;\mu\nu} \delta\sigma_{\vp\nu}^y(\omega).
\end{split}
}
Note that in the time domain, $\delta\tilde\Delta'_{\vk\mu}(t)$ and $\delta\tilde\Delta''_{\vk\mu}(t)$ do not correspond to the real and imaginary parts of $\delta\Delta_{\vk\mu}(t)$ as $V_{\vk\vp;\mu\nu}$ is a complex number. We introduce tildes in the notation as it makes the similarity to previously derived results \cite{Muller_2018, Muller_2021} more transparent. $\Delta'_{\vk\mu}(t)$ and $\Delta''_{\vk\mu}(t)$ defined below Eq.~\eqref{eq:MFH}, on the other hand, are the actual real and imaginary parts. 
\\

We focus first on the oscillations of the real part of the order parameter. Using Eq.~\eqref{eq:int_band_basis} and the relation 
\begin{equation} \label{eq:deltaOP}
	\delta \Delta_{ \vk\mu} = -\mu e^{-i\phi_{\vk}} (\delta \Delta_{s} + \mu \delta \Delta_{ \mathrm{tr}} f_{ \vk}),
\end{equation}
we can introduce another decomposition 
of the singlet and triplet components as 
\begin{equation} \label{eq:delta_decomp}
\begin{aligned}
	\delta\Delta_{s}& = \delta\tilde\Delta_{s}' + i \delta\tilde\Delta_{s}'',\\
	\delta\Delta_{tr}& = \delta\tilde\Delta_{tr}' + i \delta\tilde\Delta_{tr}''.\\
\end{aligned}
\end{equation}
Here, we have defined 
\begin{equation} \label{eq:realOP}
\begin{split}
    \delta\tilde\Delta'_s(\omega) & = -\sum_{\vp\nu} e^{i\phi_{\vp}} \left[V_s \nu +  V_m f_\vp\right]\delta\sigma_{\vp\nu}^x(\omega),\\
    \delta\tilde\Delta'_{tr}(\omega) & = -\sum_{\vp\nu} e^{i\phi_{\vp}}  \left[V_{t} f_\vp + V_m \nu \right]\delta\sigma_{\vp\nu}^x(\omega),
\end{split}
\end{equation}
and $V_{i}=V_{0}e_{i}/4$ with $i=s,tr,m$. 
The second terms on the right hand side of Eq.~\eqref{eq:delta_decomp} are
\beq{
\label{eq:imagOP}
\begin{split}
    \delta\tilde\Delta''_{s}(\omega) & = \sum_{\vp\nu} e^{i\phi_{\vp}}  \left[V_s \nu +  V_m f_\vp\right]\delta\sigma_{\vp\nu}^y(\omega),\\
    \delta\tilde\Delta''_{tr}(\omega) & = \sum_{\vp\nu} e^{i\phi_{\vp}}  \left[V_{\mathrm{tr}} f_\vp + V_m \nu \right]\delta\sigma_{\vp\nu}^y(\omega).
\end{split}
}
The consistency of the above definitions can be verified by inserting Eqs.~\eqref{eq:realOP}, \eqref{eq:imagOP}, into Eq.~\eqref{eq:delta_decomp} and further into ~\eqref{eq:deltaOP} before comparing against Eq.~\eqref{eq:general_delta_OP} with Eq.~\eqref{eq:int_band_basis}. \\

In the next step, we derive a closed set of equations for the variations of the order parameter components, as defined in Eqs.~\eqref{eq:realOP} and \eqref{eq:imagOP}. To this end, let us define
\begin{equation} \label{eq:fourComp}
  \delta \tilde{\boldsymbol{\Delta}} = \begin{pmatrix}
    \delta \tilde{ \Delta}_{s }^{ '} & \delta\tilde{ \Delta}_{s }^{''} & \delta\tilde{ \Delta}_{ \mathrm{tr}}^{'} & \delta\tilde{ \Delta}_{ \mathrm{tr} }^{ ''}  
  \end{pmatrix} ^T
\end{equation}
and
\begin{equation}
  { G}_{ \vk \mu}= e^{-i\phi_{\vp}} 
  \begin{pmatrix}
    -\mu & 0  &  -{ f}_{\vk} & 0  \\
    0 & \mu & 0 & { f}_{\vk}  \\
    0 & 0 & 0 & 0 \\
  \end{pmatrix}. 
\end{equation}
We also introduce the following interaction matrix
\begin{equation}
   V =
  \begin{pmatrix}
    V_{s} & 0 & V_{m} & 0 \\
    0 & V_{{s}} & 0 & V_{{m}} \\
    V_{m} & 0 & V_{tr} & 0 \\
    0 & V_{m} & 0 & V_{tr} \\
  \end{pmatrix} .
\end{equation}
With these definitions, Eqs.~\eqref{eq:realOP} and \eqref{eq:imagOP} can be rewritten in the form
\begin{equation} \label{eq:deltaDelta}
\begin{aligned}
  \delta \tilde{\boldsymbol{\Delta}}(\omega) &= \sum_{\vk\mu}  { V} { G}_{ \vk ,\mu}^{ \dagger } \delta  \boldsymbol{\sigma}_{ \vk\mu }(\omega) \\
  &= \sum_{\vk\mu}  { V} { G}_{ \vk ,\mu}^{ \dagger } { M}_{ \vk , \mu}(\omega) \delta \boldsymbol{ b}_{ \vk \mu}(\omega) .\\
\end{aligned}
\end{equation}
Considering the definition of $ \boldsymbol{ b}_{ \vk , \mu} $ below Eq.~\eqref{eq:MFH}, we get 
\begin{equation} \label{eq:eqForDeltaB}
  \delta \boldsymbol{ b}_{ \vk , \mu}(\omega) = { F}_{ \vk , \mu} \delta \boldsymbol{\Delta}(\omega) + \delta {b}_{ \vk \mu}^z (\omega) \boldsymbol{e}_{z},
\end{equation}
where 
$$
\begin{aligned}
F_{\vk\mu} &= \begin{pmatrix}
    -\mu \cos(\phi_{\vk}) & -\mu \sin(\phi_{\vk}) & -f_{\vk} \cos(\phi_{\vk}) & -f_{\vk}\sin(\phi_{\vk}) \\
    -\mu \sin(\phi_{\vk}) & \mu \cos(\phi_{\vk}) & -f_{\vk} \sin(\phi_{\vk}) & f_{\vk}\cos(\phi_{\vk}) \\
    0 & 0 & 0 & 0 \\
  \end{pmatrix} \\
 \end{aligned}
$$
and $\delta{\boldsymbol{\Delta}} = \begin{pmatrix}
    \delta{ \Delta}_{s }^{ '} & \delta{ \Delta}_{s }^{''} & \delta{ \Delta}_{ \mathrm{tr}}^{'} & \delta{ \Delta}_{ \mathrm{tr} }^{ ''}  
  \end{pmatrix} ^T$ is a vector of the real and imaginary parts of $\delta\Delta_{s}$ and $\delta\Delta_{tr}$. $\delta{\boldsymbol{\Delta}}$ is related to $\delta\tilde{\boldsymbol{\Delta}}$ through
\begin{equation} \label{eq:imag_OP}
  	\delta{\boldsymbol{\Delta}}(\omega)= X \delta\tilde{\boldsymbol{\Delta}}(\omega) + X^* \delta{\tilde{\boldsymbol{\Delta}}}^* (-\omega),
\end{equation}
  where $X$ is the hermitian matrix 
  $$X =\frac{1}{2} - \frac{1}{2}\mathrm{diag}( \tau^{y}, \tau^{y} ).$$
In the time domain, $\delta\boldsymbol\Delta(t)$ is clearly real as $\delta\boldsymbol\Delta(\omega)=\delta\boldsymbol\Delta(-\omega)^{*}$, enabling us to find a closed solution for $\delta\boldsymbol\Delta(\omega)$ through Eqs.~\eqref{eq:deltaDelta} and \eqref{eq:eqForDeltaB}. The result is
\begin{equation} \label{eq:finalDeltaEq}
\begin{aligned}
  \delta \boldsymbol{\Delta}(\omega) =& \, (1-\chi(\omega))^{-1} \Big[ \sum_{\vk\mu} \Gamma_{\vk\mu} {M}_{\vk\mu}(\omega)  \delta b_{\vk, \mu}^z (\omega) \boldsymbol{e}_{z} \\
  &\hspace{13mm}+ \Big(\sum_{\vk\mu} \Gamma_{\vk\mu} {M}_{\vk\mu}(-\omega)  \delta b_{\vk, \mu}^z (-\omega) \boldsymbol{e}_{z}\Big)^* \,\Big], \\ 
\end{aligned}
\end{equation}
where we defined
\begin{equation}
\begin{aligned}
  \chi( \omega) =&\sum_{ \vk \mu} \Gamma_{\vk\mu} { M}_{ \vk \mu}(\omega) { F}_{ \vk \mu} 
  + \Big( \sum_{ \vk \mu} \Gamma_{\vk\mu} { M}_{ \vk \mu}(-\omega) { F}_{ \vk \mu}\Big)^* 
\end{aligned}
\end{equation}
with
\begin{equation}
	\Gamma_{\vk\mu} = X V G_{\vk\mu}^{\dagger}.
\end{equation}
The zeros of $\mathrm{det}(1-\chi( \omega))$ yield the frequencies of the possible collective modes, while $\delta \boldsymbol{\Delta}(\omega)$ represents the amplitudes of the order parameter components. Equation~\eqref{eq:finalDeltaEq} constitutes a central result of this paper and will be used in the following section to describe the response of the spin current to an external electric field. \\

Let us briefly consider 
what happens if $\tilde{\Delta}(t)$ is real, which could occur if the Coulomb matrix elements of Eq.~\eqref{eq:int_band_basis} are real. In this case, $\tilde{\Delta}(\omega) = \tilde{\Delta}(-\omega)^{*}$, and by Eq.~\eqref{eq:imag_OP}, $\tilde{\Delta}(\omega) = \Delta(\omega)$. The previous expressions thus simplify to 
\begin{equation}
  \delta \boldsymbol{\Delta}(\omega) = (1-\chi(\omega))^{-1} \sum_{\vk\mu} \Gamma_{\vk\mu} {M}_{\vk\mu}(\omega)  \delta b_{\vk\mu}^z (\omega) \boldsymbol{e}_{z} 
\end{equation}
with
\begin{equation}
  \chi( \omega) =\sum_{ \vk \mu} \Gamma_{\vk\mu} { M}_{ \vk \mu}(\omega) { F}_{ \vk \mu},
\end{equation}
which is similar to the equations in Ref.~\cite{Muller_2021}.

\subsection{Spin current} \label{sec:C}
In this section we derive an expression for the spin current in the superconducting state. A hallmark of the Higgs mode is its signature in the THG response \cite{Murotani_2017}. Here, we derive a similar result for the second harmonic signal of the spin current.  

The spin current describes a situation in which electrons carry a flow of spin angular momentum with or without an accompanying charge current \cite{Bercioux_2015}. In materials with SOC, spin is not conserved, which prohibits a straightforward definition from a continuity equation without a source term. The standard definition employed in some of the literature (see, e.g., Refs.~\cite{Hamamoto_2017, Vorontsov_2008}) captures the phenomenon well, but predicts equilibrium DC spin currents without externally applied electric fields \cite{Rashba_2006}. 
Nevertheless, we will in the following adopt the standard definition as the DC component will not be of interest to us. The spin current is then defined as
\begin{equation} \label{eq:spinCurrDef}
\begin{aligned}
	{J}_{ij}(t) &= \sum_{\vk} \langle  \hat{j}_{\vk,ij} (t) \rangle,
\end{aligned}
\end{equation}
where the momentum-resolved expression in the spin-basis reads
$$
	\hat{j}_{\vk,ij}(t)=\frac{1}{2}  { \hat{ \psi} }_{ \vk }^\dagger  \{ \tau^{i}, v_{\vk -q\boldsymbol{A}(t),j} \}  { \hat{ \psi} }_{ \vk } ,
$$
with $v_{\vk -q\boldsymbol{A}(t),j} = q\partial_{k_j} h_{\vk-q\boldsymbol{A}(t)}$, $i,j=x,y,z$ and $ { \hat{ \psi} }_{ \vk }= \begin{pmatrix}
  { \hat{ c} }_{ \vk , \uparrow} & { \hat{ c} }_{ \vk , \downarrow} 
\end{pmatrix}^T $ \cite{Rashba_2003}.

 We will focus on the $yx$ component of the spin current and study its response to an electric field. While other nonlinear effects, such as THG, is predicated upon the emission of radiation from accelerated charges, the spin current itself does not give rise to radiation other than possibly through magnetic dipole radiation and is therefore harder to detect by the same means. However, other means of detecting the spin current exist, such as the induced precession of magnetic moments in a neighboring ferromagnet \cite{Maekawa_2013, Bercioux_2015}.

Observables can be transformed from the spin basis to the band basis with the operator $U_{\vk}$ defined in Eq.~\eqref{eq:nonintham}, 
\begin{equation}
\begin{aligned}
  \hat{ \mathcal{O}} (t) &= \sum^{}_{\vk} { \hat{ A} }_{ \vk }^{ \dagger } \mathcal{O}_{\vk}^b (t) { \hat{ A} }_{ \vk }  , 
\end{aligned}
\end{equation}
where $\mathcal{O}_{\vk}^b (t) =U_{\vk}^{ \dagger } \mathcal{O}_{\vk}(t) U_{\vk}$, with $\mathcal{O}_{\vk}(t)$ the operator in the spin basis, and $ { \hat{ A} }_{ \vk }= \begin{pmatrix}
  { \hat{ a} }_{ \vk , +} & { \hat{ a} }_{ \vk , -} 
\end{pmatrix}^T = U_{\vk}^{ \dagger } { \hat{ \psi} }_{ \vk } $.  
If we now take expectation values and approximate $ \langle  { \hat{ a} }_{ \vk ,\mu}^{ \dagger } { \hat{ a} }_{ \vk , \nu} \rangle \approx 0$ for $ \mu \neq \nu$, we arrive at the useful rewriting
\begin{equation} \label{eq:gen_obs}
\begin{aligned}
  \hat{\mathcal{O}}(t) =& \frac{ 1}{ 2} \sum^{}_{ \vk\mu } \big[  [\mathcal{O}_{\vk}^{ b}(t)]_{\mu\mu} +  [\mathcal{O}_{-\vk}^{ b} (t) ]_{\mu\mu} \big]  \Big( \hat{ \sigma}_{ \vk\mu}^{ z}(t) + \frac{ 1}{ 2}  \Big) \\
  &+ \frac{ 1}{ 2} \sum^{}_{ \vk\mu } \big[ [\mathcal{O}_{\vk}^{ b}(t)]_{\mu\mu} -  [\mathcal{O}_{-\vk}^{ b} (t) ]_{\mu\mu} \big] \Big( \hat{ \sigma}_{ \vk\mu}^{ 0}(t) - \frac{ 1}{ 2}  \Big) ,
\end{aligned}
\end{equation}
with the subscript $\mu=1(+), 2(-)$ indexing matrix elements and $\hat\sigma_{ \vk,\mu}^i$ as defined in Eq.~\eqref{eq:Pauli_operators}. The approximation $ \langle  { \hat{ a} }_{ \vk ,\mu}^{ \dagger } { \hat{ a} }_{ \vk , \nu} \rangle \approx 0$ essentially implies that the electric field only shifts the electron population within each band, but does not induce a charge transfer between the bands. In the above expression, we have used that 
\begin{equation} \label{eq:operator_sigma}
\begin{aligned}
	&{ \hat{ a} }_{ \vk\mu}^{ \dagger } { \hat{ a} }_{ \vk\mu} = \hat{ \sigma}_{ \vk\mu}^{z} + \hat{ \sigma}_{ \vk\mu}^{ 0}, \\
	&{ \hat{ a} }_{ -\vk\mu}^{ \dagger } { \hat{ a} }_{ -\vk\mu} = \hat{ \sigma}_{ \vk\mu}^{ z} - \hat{ \sigma}_{ \vk\mu}^{ 0} +1, \\
\end{aligned}
\end{equation}
which follow from Eq.~\eqref{eq:Pauli_operators} and the anti-commutation relations of the fermionic operators. Furthermore, when $E_{\vk, \mu} = E_{-\vk, \mu}$ as is the case here, $\sigma_{\vk,\mu}^0 =\frac{1}{2}$, rendering the last term in Eq.~\eqref{eq:gen_obs} zero in equilibrium. 

In order to perform the transformation to the band basis, we need to compute ${ {j}}_{\vk, yx}^{ b} ( t)= { U}_{\vk}^{ \dagger } {j}_{\vk,ij}(t) { U}_{\vk}$. Since 
\begin{equation} \label{eq:transf}
  \begin{aligned}
 	{ U}_{\vk}^{ \dagger } { \tau}^{ x} { U}_{\vk}  &= \cos(\phi_{\vk}) \tau^z + \sin(\phi_{\vk}) \tau^y ,\\
 	{ U}_{\vk}^{ \dagger } { \tau}^{ y} { U}_{\vk}  &= \sin(\phi_{\vk}) \tau^z - \cos(\phi_{\vk}) \tau^y ,\\
        { U}_{\vk}^{ \dagger } { \tau}^{ z} { U}_{\vk}  &= \tau^x ,\\
      \end{aligned}
\end{equation}
\\
where $\phi_{\vk}=\tan^{-1}(d_{\vk,y}/d_{\vk,x})$, 
we get
\begin{equation}
\begin{aligned}
   { j}_{ yx; \vk}^{ b}(t)
   =&  \, q\partial_{k_x} \epsilon^{0}_{\vk -q\boldsymbol{A}(t)} [ \sin(\phi_{\vk}) \tau^z - \cos(\phi_{\vk}) \tau^y ] \\ 
   &+ q\tau^0 \partial_{k_x} { d}_{\vk -q\boldsymbol{A}(t), y}. \\
\end{aligned}
\end{equation}
As $\delta \sigma_{\vk\mu}^0 (t)= 0$ by virtue of Eq.~\eqref{eq:EoM}, the time-dependent spin current's $\mathcal{O}(A^2)$ contributions can be written as 
\begin{widetext}
\begin{equation} \label{eq:TDSC}
\begin{aligned}
  J_{ yx}^{A^2} (t) &= \frac{ 1}{ 2}  \sum_{\vk\mu} ( [ {j}_{\vk, yx}^{ b} (t) ]_{\mu \mu} + [ {j}_{-\vk, yx}^{ b} (t) ]_{\mu\mu}  )\bigg|_{A^2 }
   \cdot ( { \sigma}_{ \vk , \mu}^{ z}(t) + \frac{ 1}{ 2}  ) \\
  &= q\sum_{\vk\mu} \delta { \sigma}_{ \vk,\mu}^{ z}(t) \Big[ \partial_{k_x} \epsilon^0_{\vk}  \sin(\phi_{\vk})  { \tau}^{ z}_{ \mu\mu} +  { \tau}^{ 0}_{ \mu\mu} \partial_{k_x} d_{\vk,y} \Big]\\
   &\hspace{3mm}+\frac{q}{2}(qA_{i}(t))(qA_{j}(t)) \sum_{\vk\mu} \Big[ \partial_{k_x} \partial_{k_i}\partial_{k_j} \epsilon^0_{\vk} \sin(\phi_{\vk})  { \tau}^{ z}_{ \mu\mu} 
  + {\tau}^{ 0}_{ \mu\mu} \partial_{k_x} \partial_{k_i}\partial_{k_j} d_{\vk, y} \Big]
  \bigg( \frac{E_{\vk \mu} - \epsilon_{\vk\mu} \mathrm{tanh}(E_{\vk\mu} \beta /2) }{2 E_{\vk \mu} } \bigg). \\
\end{aligned}
\end{equation}
\end{widetext}
Note that the expression $\big( \frac{E_{\vk \mu} - \epsilon_{\vk\mu} \mathrm{tanh}(E_{\vk\mu} \beta /2) }{2 E_{\vk \mu} } \big)$ tends to $n_{F}(\epsilon_{\vk\mu})$ in the limit $\Delta \rightarrow 0$, while the first term vanishes as $\delta { \sigma}_{ \vk,\mu}^{ z}(t)=0$ there, yielding an intuitive normal state response for the second harmonic signal.\\

Focusing on the superconducting response, we rewrite the superconducting part of Eq.~\eqref{eq:TDSC} in Fourier space,
\begin{equation} \label{eq:spinCurrFreq}
\begin{aligned}
  J_{ yx}^{A^2} ( \omega) &=  q\sum^{}_{\vk\mu} \delta \sigma_{\vk\mu}^z(\omega) 
  \Big[\partial_{k_x} \epsilon^{0}_{\vk} \sin(\phi_{\vk}) { \tau}^{ z}_{ \mu\mu} 
  + { \tau}^{ 0}_{ \mu\mu} \partial_{k_x} { d}_{ \vk, y}  \Big], \\
  \end{aligned}
\end{equation} 
which provides insights into the connection between the spin current and collective modes. In particular, the factor $\delta \sigma_{\vk\mu}^z(\omega)$ gives rise to the the Higgs mode, along with a phase mode and charge density fluctuations. For a fixed $\vk$, we may assume $ \Delta_{ \vk\mu}^{''} = 0$ without loss of generality. We then find from Eq.~\eqref{eq:BlochFT} that
\begin{equation} \label{eq:deltaz}
\begin{aligned}
	\delta \sigma_{\vk\mu}^z (\omega ) =& \frac{\tanh(\beta E_{ \vk\mu}/2) }{E_{\vk\mu}(  \omega^2 - 4E_{\vk,\mu}^2 ) }\big[ -2\epsilon_{\vk,\mu} \Delta_{\vk\mu}' \delta \Delta_{\vk\mu}' (\omega) \\
	&\quad + i\omega \Delta_{\vk\mu}' \delta \Delta_{\vk\mu}'' (\omega) + 2 \Delta_{\vk\mu}'^{2} \delta b_{\vk\mu}^z (\omega) \big].
\end{aligned}
\end{equation}
Ref.~\cite{Murotani_2017} identified a similar equation comprised of three terms corresponding to the Higgs, phase and charge density fluctuations. The last term has in other works been deemed important and in fact dominant for most electric field polarizations \cite{Cea_2016}. Hence, the divergent enhancement in nonlinear responses when $\omega\rightarrow |\Delta| $ does not necessarily originate from the dynamics of the order parameter, 
but can also come from
the fluctuations in the density response.  

The decomposition in Eq. ~\eqref{eq:deltaz} is however ill-suited for our purpose. In contrast to previous works, our decomposition into real and imaginary parts of $\Delta_{\vk\mu}$ will be inextricably linked to the momentum dependence of $e^{-i\phi_{\vk}}$ in Eq.~\eqref{eq:orderparmexpression}, so that the decomposition into Higgs, phase and charge density contribution is not straightforward. Instead, we will adopt the method of Ref.~\cite{Klein_2024}, wherein the decomposition into real and imaginary parts is performed on the momentum-independent part, $\Delta_{s} + \mu\Delta_{tr}$, of Eq.~\eqref{eq:orderparmexpression}. The charge density contribution to the spin current is then defined as the remaining terms, after the calculation of the Higgs and phase contributions.

\begin{figure*}[t]
  \includegraphics[width=0.9\textwidth]{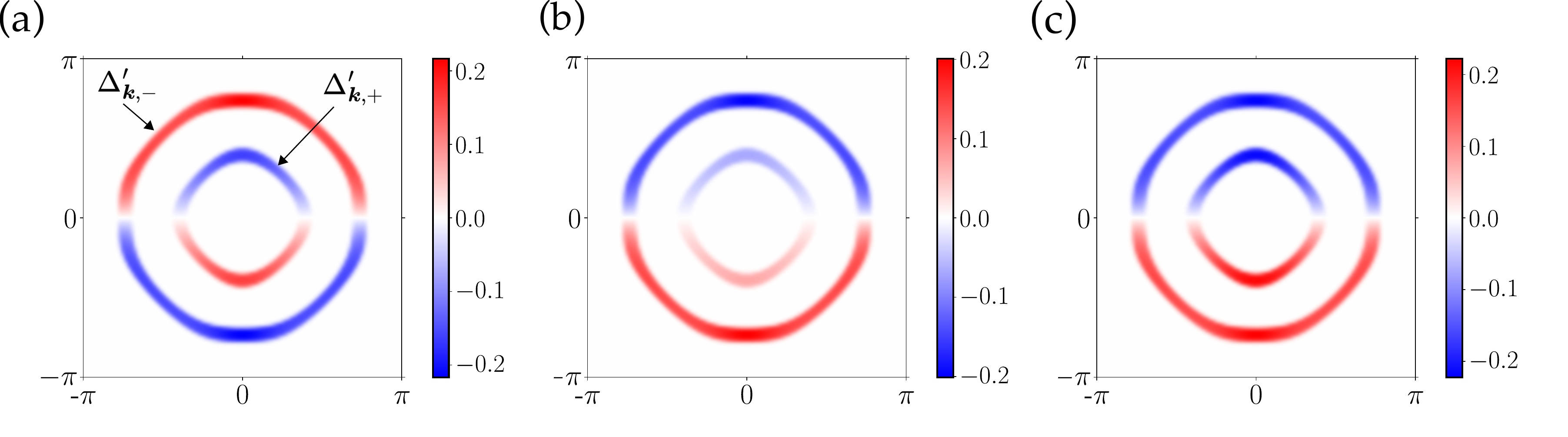} 
  \caption{ Real part of the order parameters $\Delta_{k+}$ and $\Delta_{k-}$ for $\alpha=1.2$, $V_{0}=-16$, $\beta=70$, Debye frequency $\omega_{D}=0.3$ and the following pairings: (a) $e_s=1, e_{tr}=0, e_{m}=0$, (b) $e_{s}=0.7, e_t=0.714, e_{m}=0$,  (c) $e_{s}=0, e_t=1, e_{m}=0$.}
  \label{fig:FS_real}
\end{figure*}

\subsection{Time dependent mean-field theory} \label{sec:mftheory}
To test the validity of the assumptions made in the previous sections, we consider a more general treatment which captures interband dynamics of the superconductor by means of time dependent mean-field theory. This approach also captures the effects of the self-consistent adjustment of the order parameters at each time step, as opposed to using parameters determined in equilibrium. Defining $\hat{A}_{\vk} =\big(
    { \hat{ a} }_{ \vk,+} , { \hat{ a} }_{ \vk,-} , { \hat{ a} }_{ -\vk,+}^{ \dagger } , { \hat{ a} }_{ -\vk,-}^{ \dagger } \big)^T$, and using Eq.~\eqref{eq:SOC}, we can express the time dependent mean-field Hamiltonian in Nambu space as $\hat{H}(t) = \hat{H}_{0}(t)+  \hat{ H}_{I}(t)$, where

\begin{equation} \label{eq:non_int_H}
  \begin{aligned}
  &\hat{ H}_0(t) =
  \!\sum_{ \vk} { \hat{ A} }_{ \vk}^{ \dagger } {H}_{\vk}^{0}(t){ \hat{ A} }_{ \vk} ,
\end{aligned}
\end{equation}
and
\begin{equation} \label{eq:mean_field_H}
  \hat{ H}_{I}(t) = \sum^{}_{\vk} { \hat{ A} }_{ \vk}^{ \dagger } {H}_{\vk}^{\Delta}(t)
  \hat{ A}_{ \vk}.
\end{equation}
Here, we have defined
$$
H_{\vk}^0 (t)=\frac{1}{2}\begin{pmatrix} U^{ \dagger }_{ \vk} \hat{ h}_{\vk-q\boldsymbol{A}(t)} { U}_{ \vk} & 0 \\ 
0 & -  { U}^T_{- \vk} \hat{ h}^{ T}_{-\vk-q\boldsymbol{A}(t)} { U}^{ *}_{ -\vk}  \\
\end{pmatrix}
$$
and
$$
H_{\vk}^{\Delta} (t)= \frac{1}{2} \begin{pmatrix}
    0 & { \Delta}_{ \vk}(t)  \\
    { \Delta}_{ \vk}^* (t) & 0 \\
  \end{pmatrix}
$$
with ${\Delta}_{ \vk}(t) = \mathrm{diag}( { \Delta}_{ \vk,+}(t), { \Delta}_{ \vk,-}(t) )$. (A detailed derivation of Eq.~\eqref{eq:non_int_H} is provided in Appendix~\ref{app:perturbation}.) The system's time dependent density matrix can be expressed as 
$$
  { \rho}_{\vk; \alpha, \beta} (t) =  \langle{ { \hat{ A} }_{\vk, \beta}^{ \dagger }(t)  { \hat{ A} }_{\vk, \alpha} (t) \rangle},  
$$
and obeys
\begin{equation} \label{eq:Neumann}
	i\hbar \partial_{t}{\rho}_{\vk}(t) = 2[{H}_{\vk}(t), {\rho}_{\vk}(t)],
\end{equation}
where ${H}_{\vk}(t) ={H}_{\vk}^0(t) + {H}_{\vk}^{\Delta}(t)$. The factor of two in front of the commutator stems from the fact that ${ \hat{ A} }_{ \vk}^\dagger H_{\vk}(t) { \hat{ A} }_{ \vk} = { \hat{ A} }_{ -\vk}^\dagger H_{-\vk}(t) { \hat{ A} }_{ -\vk}$ up to a constant, which also holds out of equilibrium. $\Delta_{\vk\mu}(t)$ is computed self-consistently using Eq.~\eqref{eq:EnGapEq} with the self-consistency ensured at every time step, and the unitary time evolution is computed by means of a commutator-free expansion \cite{Alvermann_2011}. After expressing observables in Nambu space in the form $\hat{\mathcal{O}}(t) = \sum_{\vk} \hat{\mathcal{O}}_{\vk}(t)$, similarly to Eq.~\eqref{eq:non_int_H}, their expectation values can be computed as 
$$
	\langle\hat{\mathcal{O}}(t) \rangle = \sum_{\vk} \mathrm{Tr}[\rho_{\vk}(t) \hat{\mathcal{O}}_{\vk}(t)].
$$
A more detailed description of the mean-field equations can be found in Appendix ~\ref{app:TDMF}. \\

Although it would be possible to perform the numerical integration directly on Eq.~\eqref{eq:EoM}, the dynamics would be different from that of Eq.~\eqref{eq:mean_field_H}. While Eq.~\eqref{eq:EoM} describes a separate set of equations for each band (indexed by $\mu$), a more realistic treatment of light-matter coupling should account for the coupling between the bands, which necessitates a description as in Eq.~\eqref{eq:non_int_H}. However, the current description also has its limitations; since the light-matter coupling only affects electrons of the same orbital character, but different spins, we neglect the dipolar coupling as described in Refs.~\cite{Schuler_2021, Li_2020}. 
This contribution becomes important when Eq.~\eqref{eq:mean_field_H} is extended to multi-orbital systems.

\section{Results}
\label{sec:Results}

\subsection{Collective modes}
We consider model~\eqref{eq:SOC} with 
\begin{equation} \label{eq:model}
	\begin{aligned}
		\epsilon_{\vk}^0 &= -2 t_{0}\big(\cos(k_{x}) + \cos(k_{y})\big), \\
		\boldsymbol{d}_{\vk}&=\alpha ( -\sin(k_{y}), \sin(k_{x}),0) ,
	\end{aligned}	
\end{equation}
where $t_{0}=1$ sets the energy scale.  The Rashba SOC coupling strength is set to $\alpha=1.2$ in order to have clearly separated Fermi surfaces.  

We begin by investigating the mixed singlet-triplet character of the order parameter within our mean-field formalism. 
In this study, we neglect contributions from the Dzyaloshinskii-Moriya interaction, i.e. $e_m = 0$.
When $e_s=1, e_{t}=0$, Eq.~\eqref{eq:EnGapEq} reduces to its singlet part only. The real part of the corresponding order parameter along the Fermi surfaces is shown in Fig.~\ref{fig:FS_real}(a) for $V_{0}=-16$ and $\beta=70$. Introducing a triplet admixture can lead to a dominant triplet part in Eq.~\eqref{eq:EnGapEq}, and to a sign difference of the order parameter between the two Fermi surfaces, as shown for $e_s=0.7, e_t=0.714$ in Fig.~\ref{fig:FS_real}(b). In Fig.~\ref{fig:FS_real}(c), we show the limit of pure triplet pairing $e_s=0, e_t=1$, which exhibits a qualitatively similar behavior. 

In order to analyze the phase mode (Leggett mode) between the two bands, it is useful to define the even-parity part of Eq.~\eqref{eq:selfcons} as 
\begin{equation} \label{eq:deltamu}
	\Delta_{\mu} = \Delta_{s} + \mu  \Delta_{tr}.
\end{equation}
Note that in our system, $f_{\vk}=1$ (which is derived in appendix~\ref{app:interaction}), which simplifies Eq.~\eqref{eq:selfcons}. In the more general case, as presented in Ref.~\cite{Klein_2024}, it is possible to define $\Delta_{\mu}$ through a factorization $\Delta_{\vk\mu}=F_{\vk}^\mu \Delta_{\mu}$. 

In addition to amplitude oscillations of the various order parameters, their phase differences have also been studied extensively in multi-band superconductors \cite{Krull_2016, Giorgianni_2019}. In a two-band superconductor, it is possible to compute the phase difference between the order parameters in the two bands, which we define as
\begin{equation} \label{eq:phase_mode}
	L(t) = \frac{1}{2\pi} [ \theta_{+}(t) - \theta_{-}(t)]
\end{equation}
with $\theta_{\mu}(t)\equiv \mathrm{arg}{\Delta_{\mu}}$. It follows from Eq.~\eqref{eq:deltamu}, however, that there can be no Leggett mode in the pure singlet case, as the relative angle is constant. The same is true for the pure triplet case, i.e., if $\Delta_{tr} = |\Delta_{tr}|e^{i\theta}$ then $L(t) = \frac{1}{2\pi} [\theta - (\theta+\pi)] = -1/2$. We thus expect the most interesting dynamics in the mixed case. 

\begin{figure}[t]
  \includegraphics[width=\linewidth]{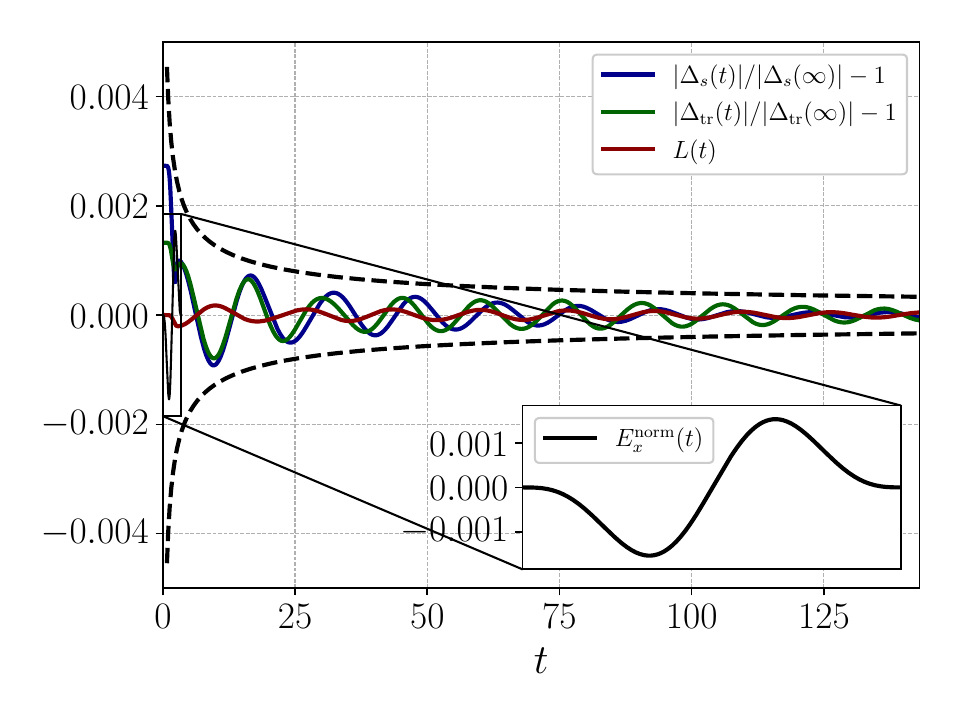} 
  \caption{Normalized oscillations of the order parameter components as well as the phase mode. The single-cycle laser pulse used to excite the system is shown in the inset. Here, $\alpha=1.2$, $E_{0}=0.2$, $\Omega=1.8$, and $\mu=-1.5$. The interaction corresponds to almost pure singlet pairing with $V_{0}=-16$ and $e_{s}=0.99$, $\beta=100$. The number of $k$-points along each dimension is $800$. 
  }
  \label{fig:coll_modes}
\end{figure}

In Fig.~\ref{fig:coll_modes}, we show the result of an excitation of a superconducting system with almost pure singlet pairing interaction, $e_\mathrm{s} = 0.99$, 
by a single cycle laser pulse with $\Omega=1.8$ and peak field strength $E_{0}=0.2$, which we indicate with an inset scaled by a normalization factor. The broad frequency spectrum of the pulse enables the activation of a vast range of excitations, and the order parameters exhibit oscillations with a frequency dictated by the magnitude of the order parameter, as well as damping. Here, we focus on the normalized singlet and triplet contributions to the order parameter, scaled by their approximate values at $t\to\infty$, to illustrate the oscillatory behavior in both the singlet and triplet channels. We note, however, that the amplitude oscillations of the triplet order parameter are relatively small. Interestingly, even for a very weak mixing between singlet and triplet pairing, oscillations in the relative phase of the order parameters, $L(t)$, remain observable. Furthermore, for the chosen parameters, and specifically for the amplitude oscillations, we observe a decay proportional to $1/\sqrt{t}$, as indicated by the black dashed lines. This property has previously been reported for conventional $s$-wave superconductors with singlet pairing interaction~\cite{Tsuji_2015}.

\begin{figure}[t]
  \includegraphics[width=\linewidth]{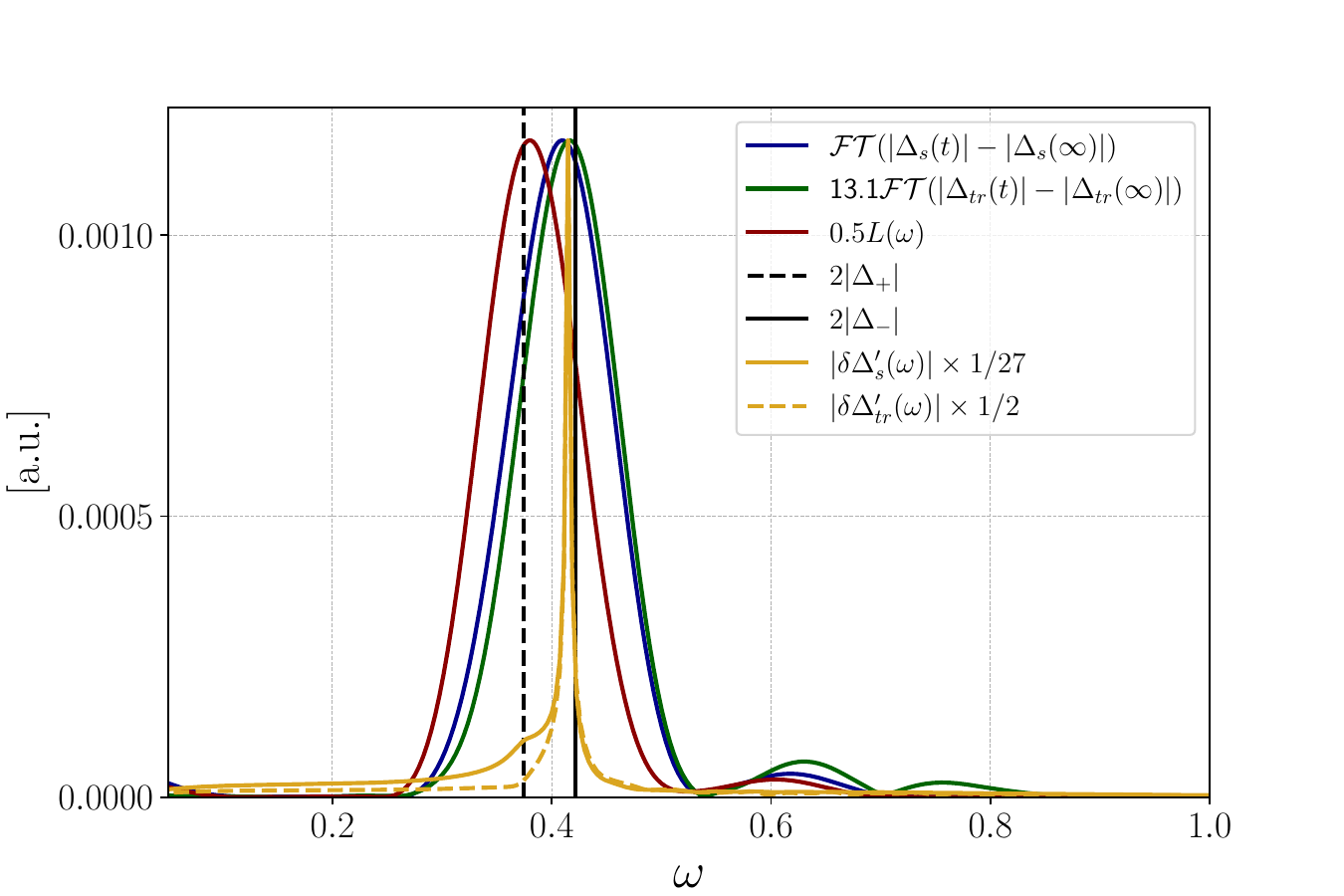} 
  \caption{Fourier transform of the collective modes computed with the time-dependent mean-field formalism with the same parameters as in Fig.~\ref{fig:coll_modes}. 
The gold colored curves are results of the analytical formulae, computed on a grid of 300 $k$-points along each axis with $\eta=0.01$.} 
  \label{fig:fourier_coll_modes}
\end{figure}

To analyze the frequency contents of the order parameters, we present in Fig.~\ref{fig:fourier_coll_modes} their Fourier transforms taken after the end of the single-cycle pulse and after applying a window function. The results of the mean-field time evolution are compared to those of the linearized equations of motion (golden colored curves). Here, we have taken $\Delta_{s}=0.198$ and $\Delta_{tr}=-0.012$ as input in Eq.~\eqref{eq:finalDeltaEq}. 

Despite the expected effective reduction of the order parameter after the field pulse in the numerical simulation, the peaks in Fig.~\ref{fig:fourier_coll_modes} align well with the frequencies of $2|\Delta_{\pm}|$. We note, however, that the triplet contribution to the amplitude fluctuations of the order parameter remains relatively weak, which leads to a correspondingly small peak in the Fourier spectrum, as discussed previously. In order to compare with the predictions of the analytical formulation, we must present a different quantity, namely $\delta\Delta_{s}'$ as derived from Eq.~\eqref{eq:finalDeltaEq}, and the comparison therefore has limitations. As for the results of the analytical calculations, the quantity $\delta\Delta_{s}'$ matches the frequency $\omega=2|\Delta_{-}|$. However, the double peak structure which is seen for the mean-field calculation is not captured except for a weak shoulder at $\omega =2|\Delta_{+}|$. Both the results of the linearized equations and the mean-field time propagation employ a Debye frequency of $\omega_{D}=0.3$ around each Fermi surface.

\subsection{Spin current}

\subsubsection{Second order response}

\begin{figure*}[t]
  \includegraphics[width=\textwidth]{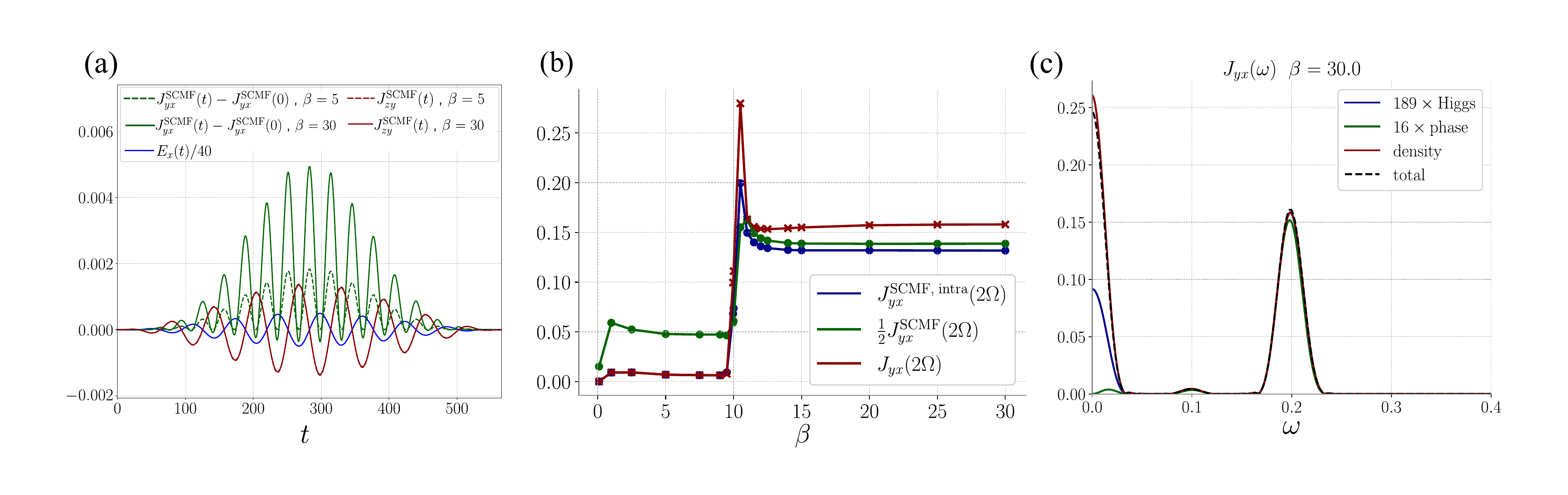} 
  \vspace{-10mm}
  \caption{ (a) 9-cycle pulse of frequency $\Omega=0.1$ and field strength $E_{0}=0.02$ and corresponding spin currents at $\beta=5$ (dashed green line) and $\beta=30$ (solid green line). (b) Comparison of the results of the mean-field dynamics and the analytical formalism ($J_{yx}$) for $e_{s}=0.99$, $e_{m}=0$. For the mean-field case, we present results with ($J^{\mathrm{SCMF}}$) and without ($J^{\mathrm{SCMF}, \; \mathrm{intra}}$) interband coupling. The results of the analytical formulation employ the same parameters, but use a broadening parameter $\eta=0.01$. (c) The Higgs, phase and density contributions [Eq.~\eqref{eq:deltaz}] to the spin current for the same parameters as in (b). The normal state contribution is not accounted for. 
  }
  \label{fig:three_panel}
\end{figure*}

In Fig.~\ref{fig:three_panel}, we demonstrate that Eqs.~\eqref{eq:spinCurrFreq} and~\eqref{eq:deltaz} hold, i.e., that the spin current shows an enhanced second order response close to the superconducting transition temperature, which we suggest to be a divergent response. We consider a pairing state that is nearly a pure singlet with $e_{s}=0.99$. Figure~\ref{fig:three_panel}(a) displays the applied laser pulse along with the resulting spin currents at two different temperatures: (i) at low temperature with $\beta=30$, and (ii) above the transition temperature with $\beta=5$. It is already evident that the spin current is larger at the lower temperature. In order to corroborate our hypothesis, we plot the results of the analytical collective modes formalism ($J_{yx}$) alongside the data from a nonequilibrium self-consistent mean-field calculation in Fig.~\ref{fig:three_panel}(b). While the latter allows to capture interband effects ($J_{yx}^\text{SCMF}$), we also present results without interband coupling ($J_{yx}^\text{SCMF, intra}$), to be more in line with the assumption made in the derivation of the analytical formulae. As input, the analytical formulation (particularly Eq.~\eqref{eq:BlochFT}) requires the magnitudes of $\Delta_{s}$, $\Delta_{tr}$, which are taken from the mean-field simulation for $J_{yx}^\text{SCMF}$. 

Qualitatively, the MF results with interband coupling ($J_{yx}^{\mathrm{SCMF}}$) and without ($J_{yx}^{\mathrm{SCMF, intra}}$), as well as $J_{yx}$ for the collective modes formalism all show a similar profile with a strong enhancement around $\beta \approx 11$. The MF result without interband coupling, $J_{yx}^{\mathrm{SCMF,intra}}$, shows the better agreement with that of the collective modes analysis, although at higher $\beta$, $J_{yx}^{\mathrm{SCMF,intra}}$ and $\frac{1}{2}J_{yx}^{\mathrm{SCMF}}$ almost coincide. The seemingly overestimated values of $J_{yx}$ compared to $J_{yx}^{\mathrm{SCMF, intra}}$ above $\beta \approx 11$ is attributed to the reduction of the order parameter as a result of the laser pulse in the mean field formalism. The peak in the spin current is similar in nature to the features found in previous studies of THG in nonequilibrium superconductors \cite{Tsuji_2015,Klein_2024, Matsunaga_2014, Cea_2016}. This suggests a similar mechanism for this peak, even though it is a second order signal in $A$ rather than a third order one.

The nonzero magnitude of the second order component (green points) of the spin current in the normal state has been established in Ref.~\cite{Hamamoto_2017}. On the other hand, the fact that the spin current \emph{without} interband pairing is much smaller in the normal state can be understood directly from Eq.~\eqref{eq:TDSC}, since the first term vanishes leaving a term proportional to $A^2$. 
The magnitude of the temperature-dependent factor in the last term of Eq.~\eqref{eq:TDSC} is bounded by $1$, rendering the whole term's contribution negligible at higher $\beta$, which supports the approximation made in Eq~\eqref{eq:spinCurrFreq}. Furthermore, while $J_{yx}(2\Omega)$ is small below $\beta\approx11$, the temperature at which the Fermi function becomes approximately constant is obviously much higher: above this temperature, any observable will be zero due to a broad electron population in the conduction band. Lastly, the kink we see at $\beta\approx 1$ for the spin current (particularly $J_{yx}^{\mathrm{SCMF}}$) likely appears because the thermal energy exceeds the Rashba SOC strength, $\alpha$, at even higher temperatures.

An interesting question regarding Fig.~\ref{fig:three_panel}(b) is the role of the Higgs contribution in the spin current response. As there is a good agreement between the analytical and numerical results for the intraband case, we may decompose the spin current according to Eq.~\eqref{eq:deltaz}. The corresponding result is shown in Fig.~\ref{fig:three_panel}(c). Similarly to previous studies on THG in superconductors \cite{Klein_2024}, we find that the contribution from the density term is the dominant one, at least for the polarization direction we have considered here. This ratio further increases away from the resonance at $\beta\approx11$. At this inverse temperature, $|\Delta_{s}| \approx 0.1$, which explains the peak at $\omega\approx 0.2$. Since previous analyses have focused on the contributions to the THG signal, let us point out that our findings for the density, Higgs and phase contributions to the THG signal are comparable to our findings for the spin current. While this appears to be in conflict with the results of Ref.~\cite{Klein_2024}, the model considered in our study is different. Our findings are more in line with the results of a $d$-wave superconductor reported in Ref.~\cite{Schwarz_2020}. However, other effects such as impurity scattering or including the effects of a retarded interaction \cite{Tsuji_2016} have been shown to greatly enhance the Higgs contribution to the THG response. \\

\subsubsection{Mean field results}

Having established the validity, as well as limitations, of our analytical theory through Fig.~\ref{fig:three_panel}(b), we now go on to analyze more realistic calculations including interband dynamics, which better capture nonlinear effects of the electric field as well as the long-time behavior. Specifically, we employ the time-dependent mean-field calculations as detailed in Sec.~\ref{sec:mftheory} and again demonstrate a strong enhancement of observables around the superconducting transition. 

\begin{figure}
  \includegraphics[width=\linewidth]{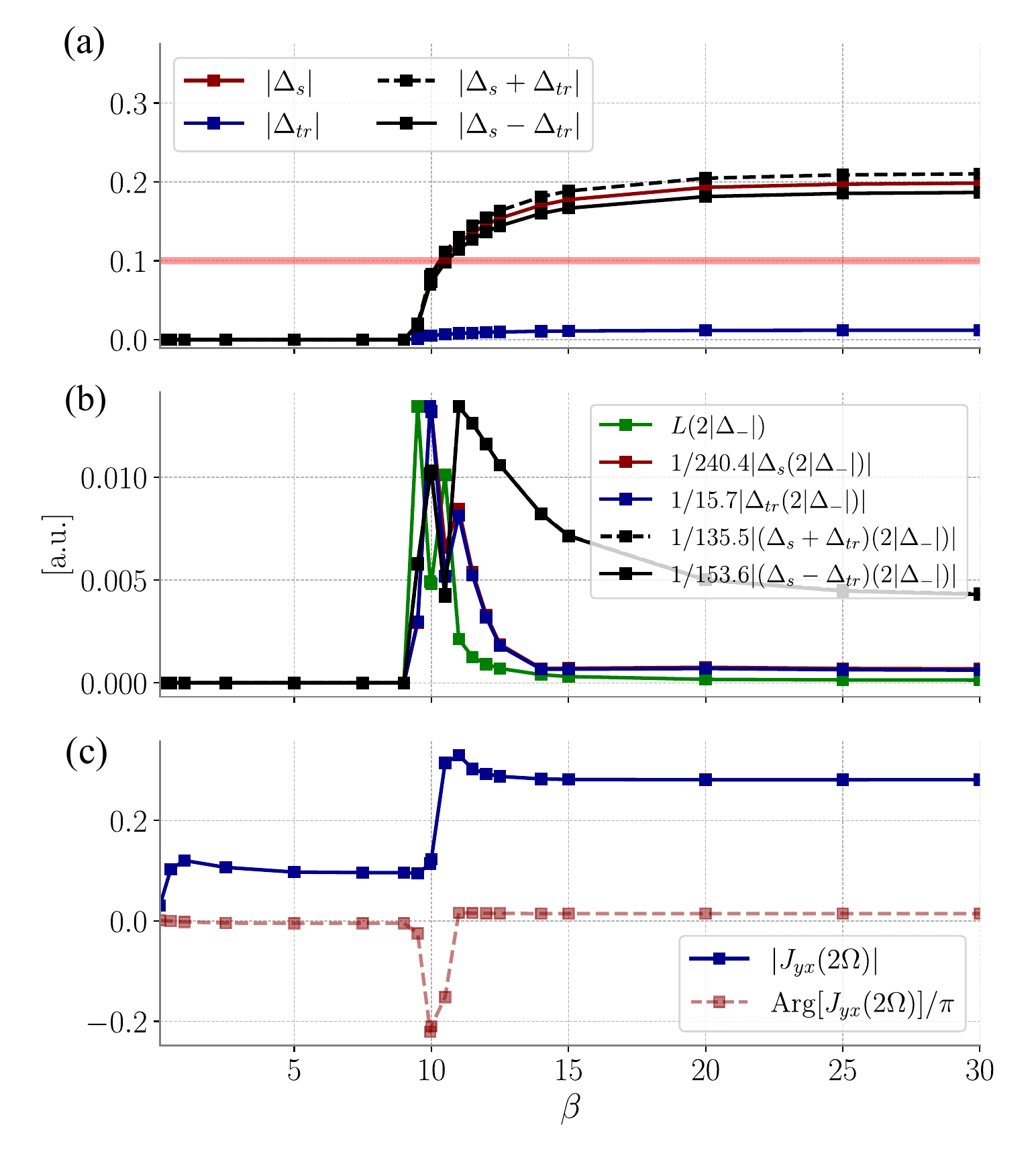}
  \caption{Time dependent mean-field results for $\alpha=1.2$, $\mu=-1.5$ with mixed singlet-triplet pairing ($V_{0}=-16$, $e_{s}=0.99$). The laser pulse has $M=9$ cycles with $E_0=0.02$ and a central frequency of $\Omega=0.1$. (a) Equilibrium order parameters and (b) magnitude of the order parameters at $\omega=2|\Delta_-|$
  for different $\beta$. (c) Fourier transform of the time dependent spin current taken at $\omega=2\Omega$. 
  }
  \label{fig:deltaSCtemp}
\end{figure}

First, we consider a system with dominant
singlet contribution to superconducting pairing,
$e_\mathrm{s} = 0.99$. As shown in  Fig.~\ref{fig:deltaSCtemp}, an enhancement of the SHG signal of the spin current is observed as the inverse temperature $\beta$ is varied across the $\Omega=|\Delta_{s}|$ resonance. The top panel displays the equilibrium order parameters as a function of the inverse temperature, $\beta$, with the red horizontal line indicating the pulse frequency. The resonance condition is determined by the crossing of the horizontal line with the magnitude of the gap function in Eq.~\eqref{eq:orderparmexpression}. We also plot the individual singlet and triplet magnitudes, $\Delta_{s}$ and $\Delta_{tr}$, in order to show the relative amplitudes of the different types of pairing. In the middle panel, we present the various order parameters as a function of $2|\Delta_{-}|$. Across the order parameter components, we see a peak splitting in several of them, which is to be expected due to the double resonance condition seen in the upper panel. Although the system exhibits a dominant singlet pairing, all the order parameters presented follow a similar profile, albeit with different magnitudes. As for the spin current, shown in the lower panel, we observe its enhancement at the Anderson pseudospin resonance, consistent with the prediction of Eq.~\eqref{eq:spinCurrFreq}. 

\begin{figure}
  \includegraphics[width=\linewidth]{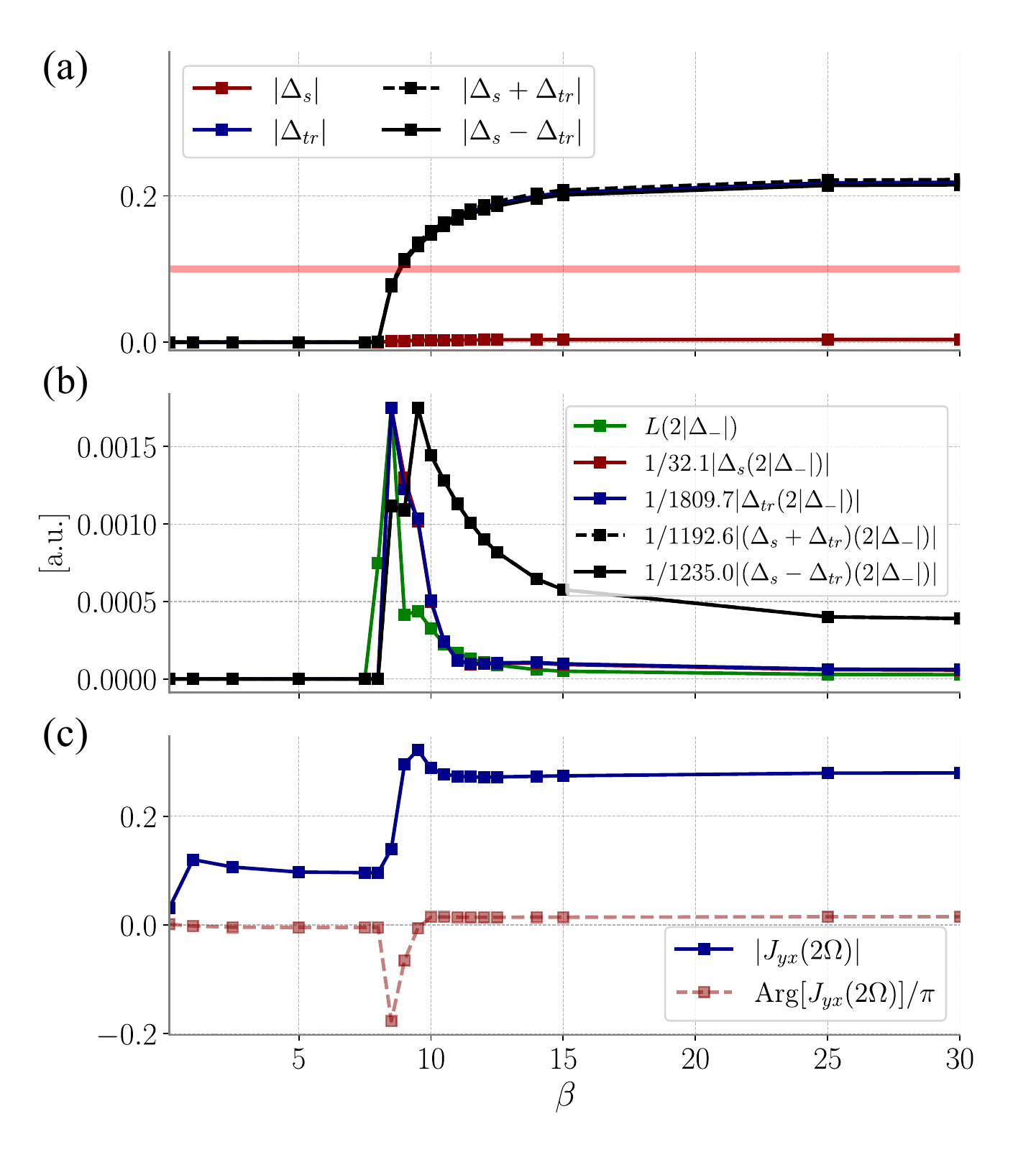}

    \caption{Time dependent mean-field results for $e_{s}=0.05$, but otherwise the same parameters and pulse as in Fig.~\ref{fig:deltaSCtemp}. (a) Equilibrium order parameters. (b) Magnitude of the order parameters at $\omega=2|\Delta_{-}|$ for different $\beta$. (c) Fourier transform of the time dependent spin current evaluated at $\omega=2\Omega$. 
    }
  \label{fig:deltaSCtemp_triplet}
\end{figure}

Next, we consider a system with dominant triplet contribution to the superconducting pairing ($e_\mathrm{s}= 0.05$). As illustrated in the upper panel of Fig.~\ref{fig:deltaSCtemp_triplet}, both superconducting order parameters of the two bands in equilibrium exhibit analogous trends as a function of the inverse temperature, shown by the black solid and dashed lines, respectively. This can be explained by the dominant triplet contribution to the superconducting pairing, which results in an almost negligible individual singlet magnitude of the order parameter (dark red line). A similar behavior is observed for the magnitude of the order parameter oscillations at $\omega = 2|\Delta_{-}|$ as a function of $\beta$, as depicted in the middle panel of Fig.~\ref{fig:deltaSCtemp_triplet}. Specifically, the oscillations of the order parameters are dominated by the triplet pairing contribution, with a peak at $\beta\approx 8$. This is reflected in an increase of the spin current at the corresponding inverse temperature, as shown in the lower panel. In general, Fig.~\ref{fig:deltaSCtemp_triplet} displays similar characteristics to Fig.~\ref{fig:deltaSCtemp}. This is not surprising in light of Fig.~\ref{fig:FS_real}: Contrary to Fig.~\ref{fig:FS_real}(b), for which $|\Delta_{\vk}|$ exhibits a greater variation along the Fermi surfaces, singlet and triplet pairing gives rise to an approximately constant profile, which in turn translates to a more well defined resonance frequency. A final point to note with respect to Figs.~\ref{fig:deltaSCtemp} and ~\ref{fig:deltaSCtemp_triplet} is that when we approach the limits of pure singlet or triplet pairing (not presented) there is no Leggett mode,  which was also seen in previous works \cite{Bittner_2015, Klein_2024} and is in line with the argument below Eq.~\eqref{eq:phase_mode}. 

\begin{figure}[t]
  \includegraphics[width=\linewidth]{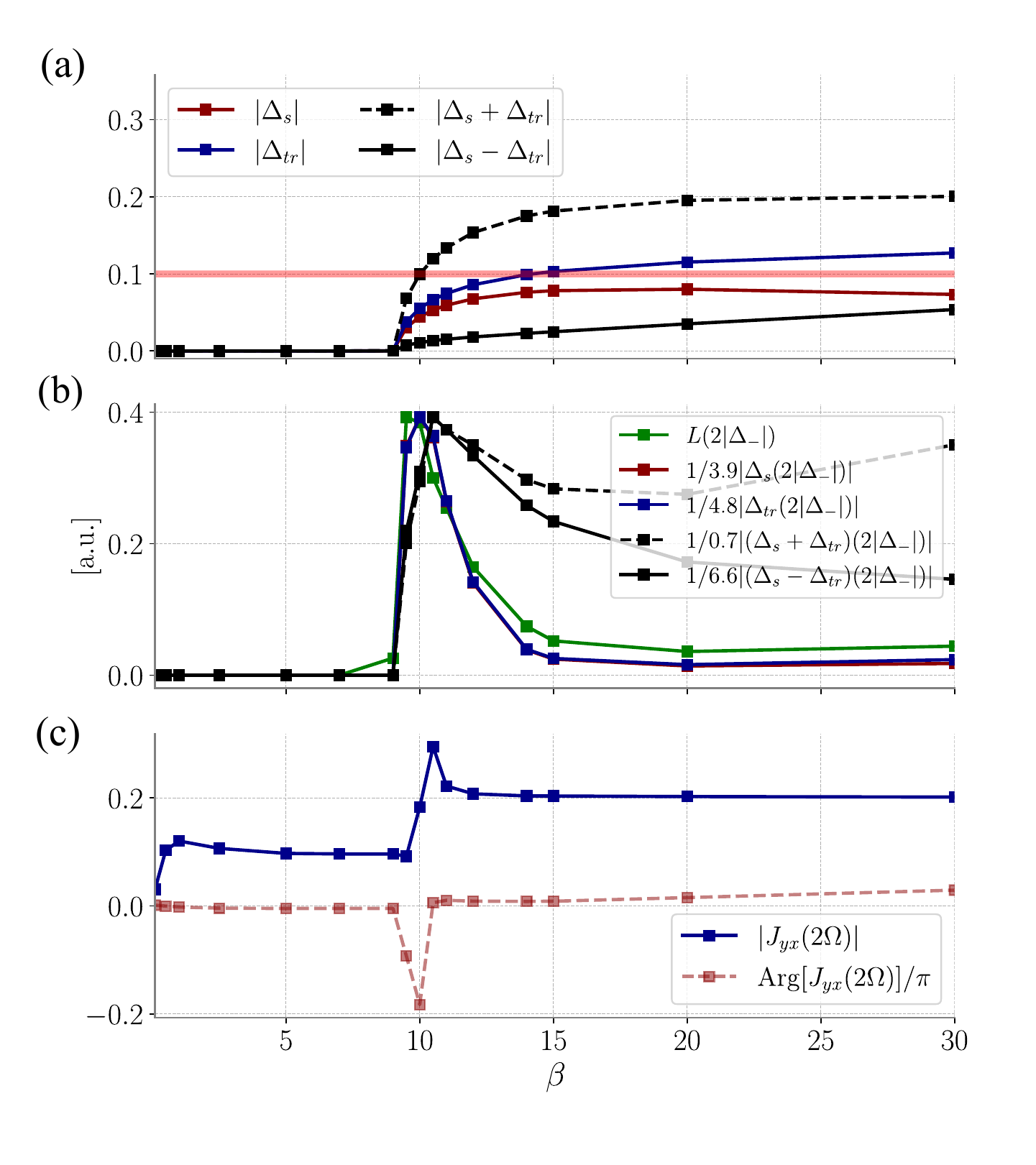} 
  \vspace{-10mm}
      \caption{Time dependent mean-field results with $e_{s}=0.7$, but otherwise the same parameters and pulse shape as in Fig.~\ref{fig:deltaSCtemp}. (a) Equilibrium order parameters and (b) magnitude of order parameters for $\omega=2|\Delta_-|$ for different $\beta$. (c) Fourier transform of the time dependent spin current evaluated at $\omega=2\Omega$.
      }
  \label{fig:deltaSCtemp_mixed}
\end{figure}

Finally, we analyze a more even singlet-triplet admixture with $e_s=0.7, e_t=0.714$ in Fig.~\ref{fig:deltaSCtemp_mixed}. Here, the equilibrium order parameter components of the two bands are split as a function of the inverse temperature, and the resonance condition with the electromagnetic field is found at both $\beta\approx10$ and $\beta\approx15$, with the second condition being a less well-defined one. The double resonance conditions is also evident in the magnitude of the order parameters at $\omega = 2|\Delta_{-}|$ as a function of $\beta$, which shows peaks around $\beta\approx10$ and large regions of enhanced order parameter components likely originating from the $\beta\approx15$ resonance. Also, for these parameters, we found that the peak of $|\Delta_\mathrm{s}-\Delta_\mathrm{tr}|$ dominates other oscillations, while the Leggett mode has the largest relative magnitude of the parameter sets considered so far. Although it is still unclear to what extent the relative contributions in the second panel is representative of the contributions to the spin current in the bottom panel, we note the simultaneous appearance of a more peaked resonance in $J_{yx}$ in the bottom panel (compared to the previous figures) and a relatively large Leggett mode, hinting at a role of the Leggett mode in enhancing the spin current resonance.

\begin{figure}[t]
\vspace{7mm}
  \includegraphics[width=\linewidth]{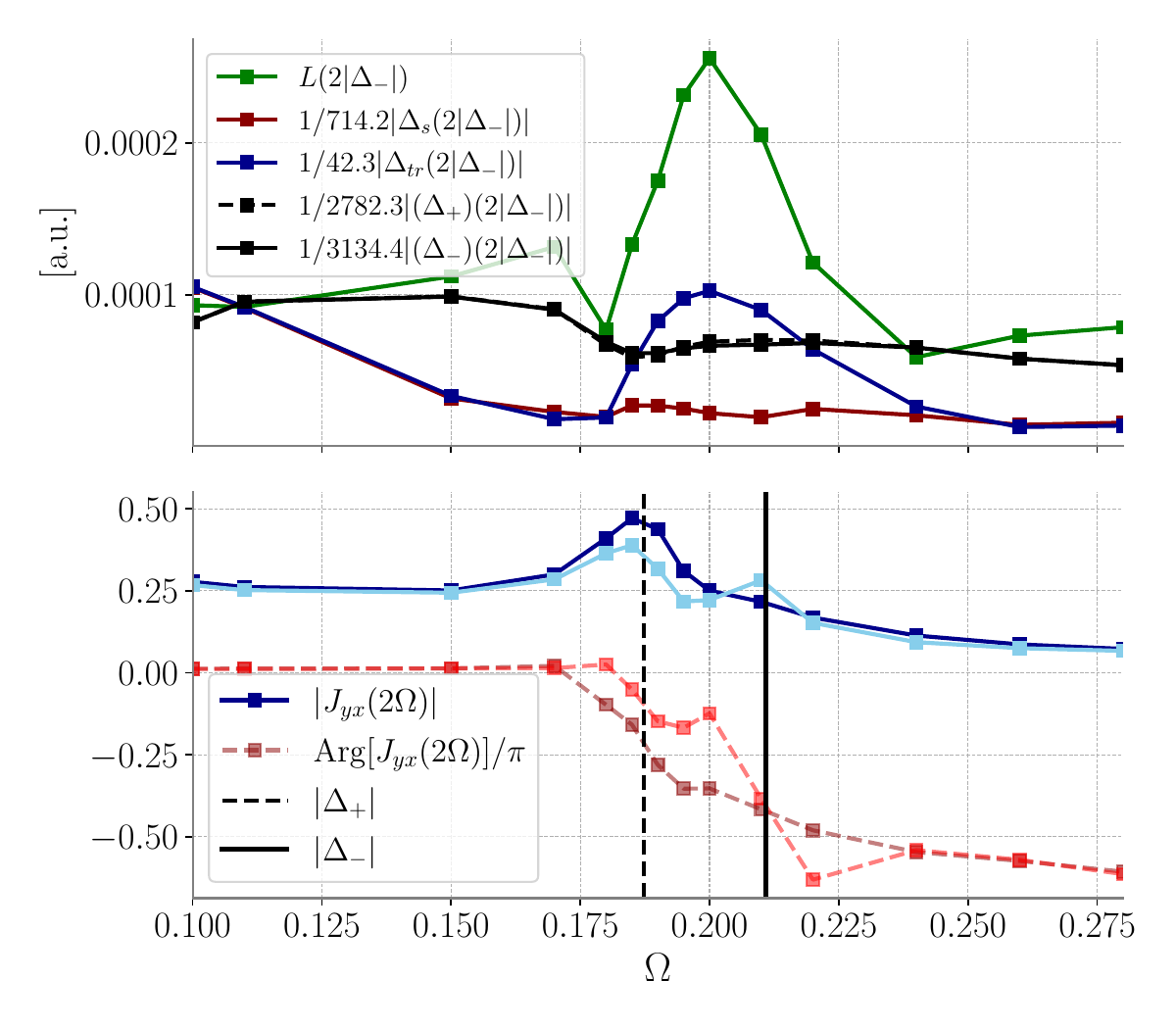}
      \caption{Time dependent mean-field results for $e_{s}=0.99$ and varying pulse frequency $\Omega$. The ratio $E_{0}/\Omega$ is kept fixed at 0.2 for all simulations ($E_{0}=0.02$ at $\Omega=0.1$) and the inverse temperature is fixed at $\beta=100$. In the lower panel, we indicate the results of the self-consistent mean field simulations in darker colors while the lighter colors represent the results without self-consistency during time evolution.}
  \label{fig:varying_freq}
\end{figure}

The resonance condition at $\Omega=|\Delta|$ is known to give rise to a phase change of $\pi/2$ in the time dependent order parameter, $\Delta(t)$, as well as in the THG susceptibility of the charge current \cite{Tsuji_2015}. It is therefore natural to ask whether such a property should also hold for the SHG signal of the spin current. While Figs.~\ref{fig:deltaSCtemp}, \ref{fig:deltaSCtemp_triplet} and \ref{fig:deltaSCtemp_mixed} show a phase change in the spin current across the superconducting transition, attributing the phase of $J_{yx}$ to the phase of any order parameter component is meaningless outside of the superconducting phase. In Fig.~\ref{fig:varying_freq}, we instead demonstrate how the magnitude and phase of the spin current evolve across the superconducting gap when the frequency of the pulse is varied but inverse temperature is held fixed at $\beta=100$. In the top panel, we show the order parameters evaluated in frequency space as a function of a fixed equilibrium order parameter value, and in the lower panel, the SHG spin current magnitude and phase. The darker colors are the results of a self consistent determination of the order parameter at each time step while the lighter colors are results of a simulation in which the order parameters are frozen to their equilibrium values. Such a distinction is made in order to more easily discern the contribution coming from order parameter dynamics and charge density fluctuations. \\

In both cases, we find a phase change of $\pi/2$ between low and high frequencies and the frequency at which it occurs is marked by a  peak in $|J_{yx}(2\Omega)|$. The enhancement at $2\Omega$ for the self-consistent case is rather modest, at least for the parameters considered. This may indicate the dominating role of charge density fluctuations -- a finding which supports the observation of Fig. ~\ref{fig:three_panel}(c). We only plot data up to $\Omega=0.3$, as for higher frequencies the overall profile of the spin current changes in such a way that a meaningful SHG phase can no longer be extracted. 

\section{Summary}
\label{sec:Summary}
In this study, we explored the nonequilibrium properties and 
nonlinear responses of non-centrosymmetric superconductors with Rashba spin-orbit coupling. 
Owing to the absence of inversion symmetry and spin orbit coupling, these superconductors not only exhibit a mixture of spin-singlet and spin-triplet pairing, but also offer an ideal platform for investigating rich spin dynamics.

Using analytical calculations based on the Anderson pseudospin formalism in combination with time-dependent mean-field simulations, we studied the signatures of collective excitations of the superconducting order parameter, i.e.  the amplitude (Higgs) and phase (Leggett) modes in the spin current response. While these modes have been shown to be resonantly enhanced in the THG signal (the lowest nonlinear excitation order of the charge current in these superconducting systems) \cite{Klein_2024, Murotani_2017}, we also identified the Higgs and Leggett modes in the second harmonic response of the spin current. In particular, an analytical form of the SHG was derived, which includes contributions from the amplitude and phase fluctuations of the order parameters, in addition to the charge density fluctuations.

Our findings indicate that charge density fluctuations provide the dominant contribution to both the spin and charge current responses. The latter is consistent with earlier studies of THG in spin-singlet superconducting systems. However, we hypothesize that the Higgs contribution to the SHG should increase in non-centrosymmetric superconductors in the dirty limit, as was recently demonstrated for THG in non-centrosymmetric superconductors \cite{Klein_2024}. This will be an interesting subject for further investigations.

To consider more realistic nonequilibrium setups and to study the signatures of the collective modes in the long-time dynamics, we performed numerical simulations, in which we additionally took into account the interband coupling. A notable observation was the substantial enhancement of the SHG signal of the spin current at the Anderson pseudospin resonance condition for different triplet-to-singlet ratios. This should be helpful for the detection of the Higgs and Leggett modes in non-centrosymmetric superconductors with low transition temperature $T_c$, since the SHG in the spin current does not produce strong heating. 

Our results open new directions for exploring spin and charge dynamics in light-driven NCS superconductors, with potential relevance for ultrafast spectroscopy and spintronic applications. Furthermore, we expect that the linearized equation of motion formalism, as presented here for the case of NCS superconductors, can be extended to multiband systems and to those with other electronic orders competing with singlet-triplet pairing. 

\appendix

\section{Derivation of the pairing interaction} \label{app:interaction}
In order to derive the pairing interaction in Eq.~\eqref{eq:orderparmexpression}, we restate Eq.~\eqref{eq:int} here:
\begin{equation}
\begin{aligned}
	\hat{H}_{I} &= \frac{1}{2}\sum_{\vk \vk'}\sum_{s_{1}s_{2} s_{1}'s_{2}'} V_{\vk \vk'; s_{1}s_{2}s_{2}'s_{1}'} \hat{c}_{\vk s_{1}}^\dagger \hat{c}_{-\vk s_{2}}^\dagger \hat{c}_{-\vk' s_{2}'} \hat{c}_{\vk' s_{1}'} \\
	&=\frac{1}{2} \sum_{\vk \vk'}\sum_{\mu\nu\gamma\delta} V_{\vk \vk'; \mu\nu\gamma\delta} \hat{a}_{\vk \mu}^\dagger \hat{a}_{-\vk \nu}^\dagger \hat{a}_{-\vk' \gamma} \hat{a}_{\vk' \delta} ,
\end{aligned}
\end{equation}
where $V_{\vk \vk'; s_{1}s_{2}s_{2}'s_{1}'}$ is defined in Eq.~\eqref{eq:interactionspinbasis} and $V_{\vk \vk'; \mu\nu\gamma\delta} $ the transformed expression in band space for which the following unitary transformation is applied
\begin{equation}
	\begin{pmatrix} \hat{c}_{\vk\uparrow} \\  \hat{c}_{\vk\downarrow} \end{pmatrix} = U_{\vk} \begin{pmatrix} \hat{a}_{\vk+} \\  \hat{a}_{\vk-} \end{pmatrix}.
\end{equation}
Here, $U_{\vk}$ is a unitary matrix which diagonalizes the Rashba Hamiltonian and is given by
\begin{equation}
	U_{\vk} = \frac{1}{\sqrt{2}} \begin{pmatrix} 1 & 1 \\ e^{i\phi_{\vk}} & -e^{i\phi_{\vk}} \end{pmatrix}, 
\end{equation}
where $\phi_{\vk} = \tan^{-1}(d_{\vk,y}/d_{\vk,x})$ and 
$e^{i\phi_{\vk}} = (d_{\vk,x} + i d_{\vk,y})/d_{\vk}$
\cite{Wolf_2020}.

Assuming only intraband pairing, 
\begin{equation}
\begin{aligned}
	\hat{H}_{I} & \approx  \frac{V_0}{8} \sum_{\vk\vk'} e_{s} R_{\vk,\mu\mu}^s  R_{\vk,\nu\nu}^{s,\dagger} \hat{a}_{\vk \mu}^\dagger \hat{a}_{-\vk \mu}^\dagger \hat{a}_{-\vk' \nu} \hat{a}_{\vk' \nu} \\
	&+ \frac{V_0}{8} \sum_{\vk\vk'} e_{tr} R_{\vk,\mu\mu}^{tr}  R_{\vk',\nu\nu}^{tr, \dagger}  \hat{a}_{\vk \mu}^\dagger \hat{a}_{-\vk \mu}^\dagger \hat{a}_{-\vk' \nu} \hat{a}_{\vk' \nu}  \\
	&+ \frac{V_0}{8} \sum_{\vk\vk'} e_{m} R_{\vk,\mu\mu}^{s} R_{\vk',\nu\nu}^{tr, \dagger} \hat{a}_{\vk \mu}^\dagger \hat{a}_{-\vk \mu}^\dagger \hat{a}_{-\vk' \nu} \hat{a}_{\vk' \nu}  \\
	&+ \frac{V_0}{8} \sum_{\vk\vk'} e_{m} R_{\vk,\mu\mu}^{tr} R_{\vk',\nu\nu}^{s, \dagger} \hat{a}_{\vk \mu}^\dagger \hat{a}_{-\vk \mu}^\dagger \hat{a}_{-\vk' \nu} \hat{a}_{\vk' \nu}  ,
\end{aligned}
\end{equation}
where 
\begin{equation}
\begin{aligned}
	R^s_{\vk} &= U^\dagger_{\vk} i\tau_{y} U^*_{-\vk} ,\\
	R^{tr}_{ \vk} &= U_{\vk}^\dagger ( \boldsymbol{g}_{\vk} \cdot \boldsymbol{\tau}) i\tau_{y} U_{-\vk}^* . 
\end{aligned}
\end{equation}
Upon performing the mean-field approximation 
$$
\begin{aligned}
	&\hat{a}_{\vk \mu}^\dagger \hat{a}_{-\vk \mu}^\dagger \hat{a}_{-\vk' \nu} \hat{a}_{\vk' \nu}  \approx \hat{a}_{\vk \mu}^\dagger \hat{a}_{-\vk \mu}^\dagger \langle \hat{a}_{-\vk' \nu} \hat{a}_{\vk' \nu} \rangle \\
	&+ \langle\hat{a}_{\vk \mu}^\dagger \hat{a}_{-\vk \mu}^\dagger \rangle \hat{a}_{-\vk' \nu} \hat{a}_{\vk' \nu} - \langle \hat{a}_{\vk \mu}^\dagger \hat{a}_{-\vk \mu}^\dagger\rangle \langle \hat{a}_{-\vk' \nu} \hat{a}_{\vk' \nu} \rangle \\
\end{aligned}
$$
we get, up to a constant term,
\begin{equation}
\begin{aligned}
	\hat{H}_{I} & \approx \frac{1}{2} \sum_{\vk\mu} \big( \Delta_{\vk\mu}^* \hat{a}_{-\vk \mu} \hat{a}_{\vk \mu} + \Delta_{\vk\mu} \hat{a}_{\vk \mu}^\dagger \hat{a}_{-\vk \mu}^\dagger \big) ,
\end{aligned}
\end{equation}
where
\begin{equation} \label{eq:deltak}
	\Delta_{\vk\mu} = R_{\vk, \mu\mu}^s \Delta_{s} + R_{\vk,\mu\mu}^{tr} \Delta_{tr}.
\end{equation}
The detailed form of the prefactors are $R_{\vk, \mu\mu}^s=-\mu e^{-i\phi_{\vk}}$ and 
$$
	R_{\vk, \mu\mu}^{tr} = -f_{\vk} e^{-i\phi_{\vk}}, 
$$
where, recalling that $\boldsymbol{d}_{\vk} = \alpha \boldsymbol{g}_{\vk}$,
$$
\begin{aligned}
	f_{\vk} &= g_{\vk, x} \cos(\phi_{\vk}) +g_{\vk,y} \sin(\phi_{\vk}) \\
	&= \sqrt{g_{\vk,x}^2 + g_{\vk,y}^2 }.
\end{aligned}
$$
This implies that in our particular case (see Eq.~\eqref{eq:model}) $f_{\vk}=1$. Furthermore, 
\begin{equation}\label{eq:opcomps}
\begin{aligned}
	\Delta_{s}=& \frac{V_{0} e_{s}}{4} \sum_{\vk\mu} R_{\vk,\mu\mu}^{s, \dagger} \langle \hat{a}_{-\vk\mu} \hat{a}_{\vk\mu} \rangle \\
	&+ \frac{V_{0} e_{m}}{4} \sum_{\vk\mu} R_{\vk,\mu\mu}^{tr, \dagger} \langle \hat{a}_{-\vk\mu} \hat{a}_{\vk\mu} \rangle , \\
\end{aligned}
\end{equation}
and 
\begin{equation}\label{eq:opcomps}
\begin{aligned}
	\Delta_{tr} =& \, \frac{V_{0} e_{tr}}{4} \sum_{\vk\mu} R_{\vk,\mu\mu}^{tr, \dagger} \langle \hat{a}_{-\vk\mu} \hat{a}_{\vk\mu} \rangle \\
	&+ \frac{V_{0} e_{m}}{4} \sum_{\vk\mu} R_{\vk, \mu\mu}^{s, \dagger} \langle \hat{a}_{-\vk\mu} \hat{a}_{\vk\mu} \rangle. \\
\end{aligned}
\end{equation}
Hence, Eq.~\eqref{eq:deltak} can be written more succinctly as 
\begin{equation}
	\Delta_{\vk\mu} = -\mu e^{-i\phi_{\vk}} (\Delta_{s} + \mu f_{\vk} \Delta_{tr}),
\end{equation}
which establishes Eq.~\eqref{eq:orderparmexpression}. 

\section{Electric field perturbation in the band basis} \label{app:perturbation}
In this section, we will establish the equivalence between the two formulations of light matter coupling in NCS superconductors considered in this paper, i.e. between Eq.~\eqref{eq:bz} and Eq.~\eqref{eq:non_int_H}. First, let us prove Eq.~\eqref{eq:non_int_H} starting from the following light matter coupling prescription, 
\begin{equation}
	\hat{H}_{0}(t)=\sum_{\vk} \begin{pmatrix} \hat{a}_{\vk,+}^\dagger & \hat{a}_{\vk,-}^\dagger \end{pmatrix} U^{ \dagger }_{ \vk} \hat{ h}_{\vk-q\boldsymbol{A}(t)} { U}_{ \vk} \begin{pmatrix} \hat{a}_{\vk,+} \\ \hat{a}_{\vk,-} \end{pmatrix}.
\end{equation}
Here, the Peierls substitution is performed in the original basis prior to a transformation to the band basis. Its equivalence to Eq.~\eqref{eq:non_int_H} is readily established by taking the average with its expression at $-\vk$ and noting that 
$$
\begin{aligned}
	&\begin{pmatrix} \hat{a}_{-\vk,+}^\dagger & \hat{a}_{-\vk,-}^\dagger \end{pmatrix} U_{-\vk}^\dagger h_{-\vk - q\boldsymbol{A}(t)} U_{-\vk} \begin{pmatrix} \hat{a}_{-\vk,+} \\ \hat{a}_{-\vk,-} \end{pmatrix} \\
	&\quad= - \begin{pmatrix} \hat{a}_{-\vk,+} & \hat{a}_{-\vk,-} \end{pmatrix} U_{-\vk}^{T} h_{-\vk - q\boldsymbol{A}(t)}^T U_{-\vk}^{*} \begin{pmatrix} \hat{a}_{-\vk,+}^\dagger \\ \hat{a}_{-\vk,-}^\dagger \end{pmatrix} \\
\end{aligned}
$$
up to a constant. The last equality results from using the anti-commutation relations for fermionic operators. Next, in order to derive Eq.~\eqref{eq:bz}, we make the approximation of only considering intraband processes. Then, using Eqs.~\eqref{eq:operator_sigma},  it is readily shown that the Hamiltonian can be written as 
\begin{widetext}
\begin{equation} \label{eq:gen_ham}
\begin{aligned}
  \hat{H}(t) =& \frac{ 1}{ 2} \sum^{}_{ \vk\mu } \bigg[  [U_{\vk}^\dagger h_{\vk - q\boldsymbol{A}(t)} U_{\vk}]_{ \mu \mu} +   [U_{-\vk}^T h_{-\vk - q\boldsymbol{A}(t)}^T U_{-\vk}^{*}]_{ \mu \mu} \bigg]  \Big( \hat{ \sigma}_{ \vk , \mu}^{ z}(t) + \frac{ 1}{ 2}  \Big) \\
  +& \frac{ 1}{ 2} \sum^{}_{ \vk\mu } \bigg[  [U_{\vk}^\dagger h_{\vk - q\boldsymbol{A}(t)} U_{\vk}]_{ \mu \mu} - [U_{-\vk}^T h_{-\vk - q\boldsymbol{A}(t)}^T U_{-\vk}^{*}]_{ \mu \mu} \bigg] \Big( \hat{ \sigma}_{ \vk , \mu}^{ 0}(t) - \frac{ 1}{ 2}  \Big).
\end{aligned}
\end{equation}
\end{widetext}
As $[U_{-\vk}^T h_{-\vk - q\boldsymbol{A}(t)}^T U_{-\vk}^{*}]_{ \mu \mu}$ is merely a rewriting of $[U_{-\vk}^\dagger h_{-\vk - q\boldsymbol{A}(t)} U_{-\vk}]_{ \mu \mu}$, Eq.~\eqref{eq:gen_ham} can be written in the form of Eq.~\eqref{eq:gen_obs} with $\mathcal{O}_{\vk}^{b}(t) = U_{\vk}^\dagger h_{\vk - q\boldsymbol{A}(t)} U_{\vk}$. 

Now, as far as the dynamics of the system in Eq.~\eqref{eq:EoM} is concerned, the only contribution from Eq.~\eqref{eq:gen_ham} is the $\hat{ \sigma}_{ \vk , \mu}^{ z}(t)$ term since $[\hat{\sigma}_{\vk\mu}^{0},\hat{\sigma}_{\vk\mu}^{i}]=0$ for $i=x,y,z$. Thus, up to a constant, the above expression effectively reduces to 
\begin{equation}
  \hat{H}(t) = \frac{ 1}{ 2} \sum^{}_{ \vk\mu } \bigg[ \tilde{\epsilon}_{\vk,q\boldsymbol{A}(t),\mu} + \tilde\epsilon_{-\vk,q\boldsymbol{A}(t),\mu} \bigg]  \hat{ \sigma}_{ \vk , \mu}^{ z}(t) 
\end{equation}
with $\tilde{\epsilon}_{\vk, \boldsymbol{p},\mu} = [U_{\vk}^\dagger h_{\vk - \boldsymbol{p}} U_{\vk}]_{\mu\mu}$ which establishes Eq.~\eqref{eq:bz}.

By a straightforward calculation, one can also show that the linear-in-$A$ term of $[U_{\vk}^{\dagger} h_{\vk - q\boldsymbol{A}(t)} U_{\vk}]_{ \mu \mu} + [U_{-\vk}^{\dagger} h_{-\vk - q\boldsymbol{A}(t)} U_{-\vk}]_{ \mu \mu}$ vanishes for the Rashba model, in line with the well known fact that the Higgs mode only couples to the quadratic order of the vector potential. \cite{Tsuji_2015} 

\section{Charge current}\label{app:charge}

Equation~\eqref{eq:spinCurrFreq}, predicts a divergent spin current at the frequency $\omega=2\Omega$. In this section, we show that the divergent THG response, which was demonstrated in Ref.~\cite{Tsuji_2015}, readily follows from our derivations as well. To that end, let us consider the $x$-component of the charge current, with $\boldsymbol{v}^b_{\vk-q\boldsymbol{A}(t)} = U^\dagger_{\vk} \boldsymbol{v}_{\vk-q\boldsymbol{A}(t)} U_{\vk}$ where $\boldsymbol{v}_{ \vk -q\boldsymbol{A}(t)} = q\nabla_{\vk} h_{\vk-q\boldsymbol{A}(t)}$. Using Eq.~\eqref{eq:gen_obs} and keeping only interband contributions, we find that the third order contributions from the vector potential are 
\begin{equation} \label{eq:deltaV}
\begin{aligned}
	J_{x}^{A^3}(t)\approx &qA_{j}(t) \sum_{ \vk, \mu} \delta { \sigma}_{ \vk,\mu}^{ z}(t) \big[ \partial_{k_{x}}\partial_{k_{j}}\epsilon^{0}_{\vk} \tau^{0}_{\mu\mu} \\
	& (\partial_{k_{x}}\partial_{k_{j}}d_{\vk, y}) \sin(\phi_{\vk})  \tau^{z}_{\mu\mu} \big],
\end{aligned}
\end{equation}
which is a THG response, as shown in previous works \cite{Tsuji_2015}. 

By the convolution theorem, we may write 
\begin{equation} \label{eq:THG_charge}
\begin{aligned}
	& J_{x}^{A^3} (\omega) = \sum_{ \vk\mu} \int_{-\infty}^{\infty} d\omega' \delta \sigma_{ \vk\mu}^{z} (\omega') A_{j}(\omega - \omega') \\
	&\quad \times q\Big[ \partial_{k_{x}}\partial_{k_{j}} \epsilon^{0}_{\vk} \tau^{0}_{\mu\mu}+  (\partial_{k_{x}}\partial_{k_{j}}d_{\vk,y}) \sin(\phi_{\vk})  \tau^{z}_{\mu\mu}  \Big],
\end{aligned}
\end{equation}
which for a monochromatic pulse, $A_{x}(\omega) \equiv A_{0} \delta_{}(\omega - \Omega) + A_{0}^* \delta_{}(\omega + \Omega)$, will give a resonant enhancement at $\omega=\Omega$ as well as $\omega=3\Omega$, with both peaks supported by the divergent enhancement of $\delta \sigma^z_{ \vk\mu}(\omega')$ at $\omega=2 E_{ \vk\mu}$. \\

\section{Time dependent mean-field theory} \label{app:TDMF}
Using the notation of Sec.~\ref{sec:mftheory} we here derive the equation of motion for the many-body density matrix in the mean-field formalism. We let 
$\hat{H}(t) = \sum_{\vk \in \mathrm{BZ}}\sum_{\mu\nu} \hat{A}_{\vk,\mu}^\dagger H_{\vk, \mu\nu}(t) \hat{A}_{\vk,\nu} $ and $\rho_{\vk;\mu\nu}(t) = \langle \hat{A}_{\vk,\nu}^\dagger \hat{A}_{\vk,\mu} \rangle(t) =  \mathrm{Tr}[\hat{\rho}(t) \hat{A}_{\vk,\nu}^\dagger \hat{A}_{\vk,\mu} ] $ for $\vk \in \mathrm{BZ}$ and note that all $\hat{A}_{\vk\mu}$ in  $\hat{A}_{\vk}  = \big(
    { \hat{ a} }_{ \vk,+} , { \hat{ a} }_{ \vk,-} , { \hat{ a} }_{ -\vk,+}^{ \dagger } , { \hat{ a} }_{ -\vk,-}^{ \dagger } \big)^T$ obey standard anti-commutation relations. 

Before proceeding, let us note that due to the presence of operators for $\vk$ and $-\vk$ in $\hat{H}(t)$, evaluating the equation of motion requires some care. In particular 
$$
\begin{aligned}
	i\hbar \partial_{t}\rho_{\vk;\mu\nu}(t) &= i\hbar \mathrm{Tr} [\partial_{t} \hat{\rho}(t) \hat{A}_{\vk,\nu}^\dagger \hat{A}_{\vk,\mu} ]\\
	&= \mathrm{Tr} [ [ \hat{H}(t), \hat{\rho}(t)] \hat{A}_{\vk,\nu}^\dagger \hat{A}_{\vk,\mu}  ] \\
	&= \mathrm{Tr} [ \hat{\rho}(t) [ \hat{A}_{\vk,\nu}^\dagger \hat{A}_{\vk,\mu},\hat{H}(t) ]] \\
	&= \mathrm{Tr} [ \hat{\rho}(t) [ \hat{A}_{\vk,\nu}^\dagger \hat{A}_{\vk,\mu},\hat{H}_{\vk}(t) + \hat{H}_{-\vk}(t) ]] \\
	&= 2\sum_{\lambda} ( H_{\vk, \mu\lambda}(t) \rho_{\vk; \lambda\nu} - \rho_{\vk; \mu\lambda} H_{\vk, \lambda\nu}(t) ) \\
	&= 2[H_{\vk}(t) ,\rho_{\vk}(t)]_{\mu\nu}, \\
\end{aligned}
$$
where we have used the cyclicity of the trace and the identity $\mathrm{Tr}[[A,B]C] = \mathrm{Tr}[B[C,A]]$ 
to go from the second to the third line. This establishes Eq.~\eqref{eq:Neumann}. In equilibrium,
$$
	\rho_{\vk;\mu\nu}= \langle  \hat{A}_{\vk,\nu}^\dagger \hat{A}_{\vk,\mu} \rangle = \delta_{\mu\nu} n_{F}(2D_{\vk,\mu\mu}),
$$
with $D_{\vk}$ the diagonal matrix of eigenvalues of ${H}_{\vk}$ and $n_{F}(\epsilon_{\vk\mu})$ the Fermi-Dirac distribution. \\

\begin{acknowledgments}
This work was supported by ERC Consolidator Grant No. 724103. 
The numerical calculations have been performed on the Beo5 cluster at the University of Fribourg. We thank D. Einzel and N. Tsuji for fruitful discussions.
\end{acknowledgments}

\bibliography{bibliography.bib}

\end{document}